\documentclass[lettersize,journal]{IEEEtran}
\usepackage{amsmath,amsfonts}
\usepackage{amssymb}
\usepackage{algorithmic}
\usepackage{array}
\usepackage{subcaption}
\usepackage{textcomp}
\usepackage{stfloats}
\usepackage{url}
\usepackage{verbatim}
\usepackage{graphicx}
\usepackage{cite}
\usepackage{epstopdf}
\usepackage[linesnumbered, ruled]{algorithm2e}
\usepackage{color}
\usepackage[table]{xcolor}
\usepackage{multirow}
\usepackage{makecell}
\usepackage{threeparttable}
\usepackage{booktabs}
\usepackage{pifont}
\usepackage{csquotes}
\usepackage{diagbox}
\usepackage{hhline}

\begin{document}

\title{LLM-Aided Joint Secrecy Precoding and Trajectory for RSMA-Based Heterogeneous UAV Networks}

\author{Lijie Zheng, Ji He,~\IEEEmembership{Member,~IEEE}, Shih Yu Chang,~\IEEEmembership{Senior Member,~IEEE}, and Yulong Shen,~\IEEEmembership{Member,~IEEE}

\thanks{L. Zheng, J. He, and Y. Shen are with the School of Computer Science and Technology, Xidian University, Xi'an, 710071 China (e-mail: lijzheng@stu.xidian.edu.cn; jihe@xidian.edu.cn; ylshen@mail.xidian.edu.cn).}
\thanks{S.Y. Chang is with the Department of Applied Data Science, San Jose State University, San Jose, CA, U. S. A. (e-mail: shihyu.chang@sjsu.edu).}
}



\maketitle

\begin{abstract}
This paper investigates secure communications in rate-splitting multiple access (RSMA) enabled heterogeneous UAV networks, where multiple UAVs collaboratively serve ground terminals in the presence of eavesdroppers. By jointly considering secrecy rate maximization and propulsion energy consumption minimization, we formulate a multi-objective optimization problem involving UAV trajectory design, service association, power allocation, and secrecy precoding under mobility, collision-avoidance, service-capacity, and communication constraints. The formulated problem is highly non-convex due to the coupling among UAV trajectories, RSMA transmission variables, and secrecy constraints. To address the resulting non-convex and highly coupled optimization problem, we propose a hierarchical optimization framework. The inner layer uses a semidefinite relaxation (SDR)-based S2DC algorithm combining penalty functions and difference-of-convex (D.C.) programming to solve the secrecy precoding problem with fixed UAV positions. The outer layer introduces a Large Language Model (LLM)-guided heuristic multi-agent reinforcement learning approach (LLM-HeMARL) for trajectory optimization. LLM-HeMARL efficiently incorporates LLM-generated expert heuristic policy, enabling UAVs to learn energy-aware, security-driven trajectories without the inference overhead of real-time LLM calls. The simulation results show that our method outperforms existing baselines in secrecy rate and energy efficiency, with consistent robustness across varying UAV swarm sizes and random seeds.
\end{abstract}


\section{Introduction}

\IEEEPARstart{W}{ITH} the rapid advancement of 6G technology, unmanned aerial vehicles (UAVs) have increasingly become a critical component of modern communication infrastructure, owing to their high mobility, strong scalability, and the provision of reliable line-of-sight (LoS) links~\cite{geraci2022will},~\cite{mozaffari2019tutorial}. However, the broadcast nature of wireless channels over LoS links makes UAV communications more susceptible to eavesdropping and jamming attacks compared to traditional terrestrial networks, which poses significant security and privacy threats~\cite{bai2022uav},~\cite{zhang2024covert},~\cite{zheng2025rsma},~\cite{bastami2021physical}. As deployment scenarios grow in complexity, collaborative networks composed of heterogeneous UAVs, i.e., heterogeneous UAV networks (HetUAVNs), are increasingly becoming the dominant paradigm in modern applications~\cite{li2024maximizing}. Usually, due to differences in payload capacity and computing resources, UAVs in these networks often exhibit different \emph{coverage range} and \emph{service capacity}. Although UAV heterogeneity enhances network functionality and environmental adaptability, it also introduces unique and formidable challenges in the realm of physical layer security (PLS).

On the one hand, UAVs equipped with high payload capacities and strong computing power typically offer extensive coverage range and substantial service capabilities. However, these advantages come at the cost of increased exposure to potential eavesdroppers (Eves) during flight, which significantly reduces the confidentiality of the system. Therefore, secure communication must be ensured through complex trajectory planning and robust precoding design. On the other hand, for UAVs with lower payloads and limited computing power, their smaller coverage range reduces some security risks. Nevertheless, they are extremely sensitive to the energy consumption of the propulsion system and place higher demands on the performance of the algorithm. Therefore, in HetUAVNs, enhancing system secrecy and minimizing the propulsion energy consumption of the entire fleet constitute two conflicting core optimization goals. In order to achieve the overall optimal system performance under the constraints caused by this heterogeneity, it becomes crucial to carefully and collaboratively design the flight trajectories and precoding strategies of the UAVs to achieve a delicate balance between enhancing system secrecy and minimizing the overall flight propulsion energy consumption of the UAV swarm.

For this highly dynamic and strongly coupled multi-objective trade-off problem, conventional optimization methods typically rely on relaxation, approximation, or heuristic search to decouple interdependent variables~\cite{li2023multi},~\cite{li2023multi2},~\cite{10643301}. Although these methods are useful for security-energy optimization, their search randomness, computational cost, and homogeneous-network assumptions limit their applicability to HetUAVNs. Deep reinforcement learning (DRL) offers a promising alternative by enabling adaptive decision-making in dynamic environments. A substantial body of research has applied DRL to solve multi-objective optimization (MOO) problems in wireless systems \cite{zhang2024multi}, \cite{song2022evolutionary}, \cite{li2024collaborative}. However, existing DRL frameworks are not directly applicable to heterogeneous UAV network environments. Specifically, the lack of effective experience sharing among heterogeneous UAVs results in poor sample efficiency. Due to the differences in mission objectives caused by UAV capabilities, common techniques for accelerating convergence and enhancing stability (such as parameter sharing) become ineffective. These severe challenges make it difficult for UAVs to determine optimal coverage areas in accordance with their heterogeneous characteristics. Moreover, blind exploration further aggravates the issues of training instability and slow convergence, hindering the effectiveness of the learning process.

Recently, owing to the remarkable natural language understanding and mathematical reasoning capabilities of LLMs, a growing body of work has introduced them into wireless optimization, including LLM-assisted MOO, convex optimization, and DRL guidance~\cite{li2025large},~\cite{li2025joint},~\cite{li2025llm}. These works indicate that LLMs can provide useful semantic reasoning and expert knowledge for complex wireless systems. This motivates our exploration of applying LLMs to heterogeneity-aware trajectory optimization in HetUAVNs. We envision that these semantic capabilities can be leveraged to generate expert-level guidance that fully accounts for the capability distinctions among heterogeneous UAVs, directing agents toward trajectories that balance both security and energy efficiency. To validate this intuition, a preliminary experiment (detailed in Section~\ref{sec:discussion}) demonstrates that even a direct LLM planner, without domain-specific training, naturally produces heterogeneity-aware trajectories, confirming the potential of LLMs for this task. However, deploying LLMs directly within the control loop proves impractical, as their high inference latency conflicts with the real-time demands of communication systems.

\begin{table*}[t]
	\centering
	\caption{\textsc{The Difference Between Our Work and The Existing Works.}}
	\label{tab:work-difference}
	\renewcommand{\arraystretch}{1.1}
	\begin{tabular}{|cc|>{\columncolor{gray!20}}c|c|c|c|c|c|c|c|c|c|c|c|c|c|}
		\hline
		\specialrule{.05em}{0pt} {.00ex}
		\multicolumn{2}{|c|}{Ref}                                                                                              & Ours & {\cite{bai2022uav}} & {\cite{zhang2024covert}} & {\cite{zheng2025rsma}} & {\cite{bastami2021physical}} & {\cite{li2024maximizing}} & {\cite{li2023multi}} & {\cite{li2023multi2}} & {\cite{10643301}} & {\cite{zhang2024multi}} & {\cite{song2022evolutionary}} & {\cite{li2025large}} & {\cite{li2025joint}} & {\cite{li2025llm}} \\ \hline
		\multicolumn{2}{|c|}{UAV Heterogeneity}                                                                                & \ding{51}    &         &         &          &          & \ding{51}       &          &          &          &          &          &          &          &          \\ \hline
		
		\multicolumn{1}{|c|}{\multirow{2}{*}{Objective}}                                                        & Security     & \ding{51}    & \ding{51}       & \ding{51}       & \ding{51}        & \ding{51}        &          & \ding{51}        & \ding{51}        &          & \ding{51}        & \ding{51}        &          &          &          \\ \hhline{~|*{15}{-}|}
		
		\multicolumn{1}{|c|}{}                                                                                  & \begin{tabular}[c]{@{}c@{}} Energy \\ Consumption \end{tabular}       & \ding{51}    &         &          &          &          &          & \ding{51}       & \ding{51}        & \ding{51}        & \ding{51}        & \ding{51}        &          &          &          \\ \hline
		
		\multicolumn{1}{|c|}{\multirow{2}{*}{\begin{tabular}[c]{@{}c@{}}Optimization \\ Approach\end{tabular}}} & Coupled     &        &         & \ding{51}       & \ding{51}        &          &          & \ding{51}       & \ding{51}        &          & \ding{51}        & \ding{51}        & \ding{51}        &          & \ding{51}        \\ \hhline{~|*{15}{-}|}
		
		\multicolumn{1}{|c|}{}                                                                                  & Hierarchical & \ding{51}    & \ding{51}       &          &          & \ding{51}        &          &          &          & \ding{51}        &          &          &          & \ding{51}        &          \\ \hline
		\multicolumn{2}{|c|}{LLM}                                                                                              & \ding{51}    &         &         &         &          &          &          &          &          &          &          & \ding{51}        & \ding{51}        & \ding{51}        \\
		\hline
		\specialrule{.05em}{0pt} {.00ex}
	\end{tabular}
\end{table*}

To address the above challenges in solving the security-energy tradeoff in RSMA-based HetUAVNs, this paper innovatively proposes a hierarchical optimization framework that integrates LLM guidance with multi-agent reinforcement learning (MARL). To the best of our knowledge, the security problem of RSMA-based HetUAVNs is still an unexplored area. The positioning of this work relative to representative studies can be clearly seen from Table~\ref{tab:work-difference}. Therefore, the main contributions of this paper are summarized as follows:
\begin{itemize}	
\item We investigate a realistic HetUAVN where UAVs with distinct coverage ranges and service capacities cooperatively serve GTs under multiple Eves. To trade off secrecy and energy efficiency, we formulate a MOO problem maximizing the secrecy rate while minimizing UAV propulsion energy, and propose a hierarchical framework that jointly designs UAV trajectories and secrecy precoding for this non-convex and coupled problem.

\item For the inner layer, we recast the problem as a secrecy precoding subproblem under fixed UAV positions, decoupling UAV motion from communication variables. To handle the non-convex constraints induced by Eves, we develop the S2DC algorithm based on SDR, exact penalty, and D.C. iteration to maximize the secrecy rate.

\item For the outer-layer trajectory optimization, we propose an LLM-driven heuristic MARL (LLM-HeMARL) method that integrates LLM-generated expert policies into MARL, guiding heterogeneity-aware trajectory learning and reducing blind exploration to improve convergence. Notably, the LLM does not participate in real-time decision-making. Instead, a combination of offline and online RL distills its expert policy into fast policies that meet the stringent latency requirements of wireless systems.

\item Extensive simulations verify the effectiveness of the proposed method in HetUAVNs. Experiments under different random seeds and UAV swarm sizes demonstrate its convergence improvement, robustness, and scalability.
\end{itemize}

The remainder of this paper is structured as follows. Section \uppercase\expandafter{\romannumeral2} introduces the system model and formulates the MOO problem. Section \uppercase\expandafter{\romannumeral3} presents the S2DC precoding design for fixed UAV positions. Section \uppercase\expandafter{\romannumeral4} describes trajectory optimization using LLM-driven heuristic MARL. Section \uppercase\expandafter{\romannumeral5} reports the simulation results, and Section \uppercase\expandafter{\romannumeral6} concludes the paper.

\emph{Notation:} $\|\mathbf{x} \|_2$ denotes the $L_2$-norm of a vector $\mathbf{x}$. $(\cdot)^H$ denotes the conjugate transpose operator. $|\cdot|$ denotes the absolute value operator.  A complex Gaussian random variable $x$ with zero mean and variance $\sigma^2$ is denoted by $x \in \mathcal{CN}(0, \sigma^2)$. $\mathbb{C}^{M \times N}$ represents the set of complex-valued $M \times N$ matrices. $\triangleq$ denotes definition, and $\text{Tr}(\cdot)$ denotes the matrix trace. $\nabla$ and $\langle\rangle$ are gradient and scalar product, respectively. $\mathbf{1}[\cdot]$ denotes the indicator function, equal to $1$ if the condition holds and $0$ otherwise.

\begin{figure}[!t]
	\centering
	\includegraphics[width=0.5\textwidth]{"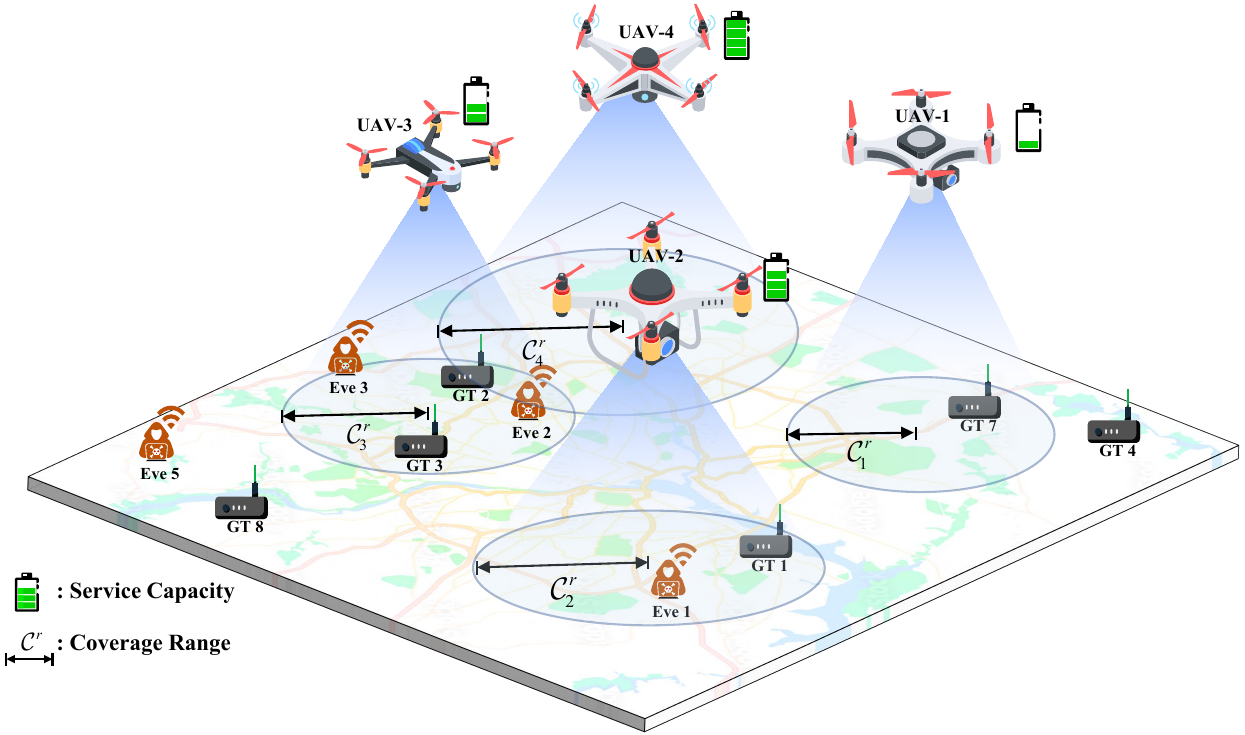"}
	\caption{Illustration of RSMA-enabled HetUAVNs}
	\label{Fig:Network_model}
\end{figure}

\section{System Model and Problem Formulation}
As illustrated in Fig.~\ref{Fig:Network_model}, we consider an RSMA-enabled multi-UAV network, which consists of $N_{\mathcal{K}}$ heterogeneous UAVs with varying payload and computing capabilities, denoted by the set $\mathcal{K} = \{1, 2, \ldots, N_{\mathcal{K}}\}$, $N_{\mathcal{I}}$ stationary ground terminals (GTs) indexed by the set $\mathcal{I}=\{1, 2, \ldots, N_{\mathcal{I}}\}$, and $N_{\mathcal{E}}$ Eves represented by the set $\mathcal{E} = \{1, 2, \ldots, N_{\mathcal{E}}\}$. Specifically, $N_{\mathcal{K}}$ UAVs flying at a fixed altitude $H_{\text{UAV}}$, each equipped with $M$ antennas, simultaneously provide downlink communication services to single-antenna GTs within an area of size $D \times D$ in the presence of single-antenna Eves. The entire duration of service is evenly discretized into $N_{\mathcal{T}}$ consecutive time slots of length $\Delta t$, denoted as $\mathcal{T}=\{1,2,\ldots, N_{\mathcal{T}}\}$. In any given time slot, the position of UAV $k$ is denoted by $u_k(t) = [x_k(t), y_k(t), H_{\text{UAV}}]$, where $x_k(t) \in [0, D]$ and $y_k(t) \in [0, D]$, $\forall k \in \mathcal{K}, t \in \mathcal{T}$. Similarly, the positions of GT $i$ and Eve $e$ are represented by $u_i = [x_i, y_i, 0]$ and $u_e = [x_e, y_e, 0]$, respectively. The length of each time slot is assumed to be sufficiently small so that the positions of UAVs and the CSI remain approximately unchanged.

\subsection{UAV Movement and Energy Consumption Models}
In the time slot $t$, UAV $k$ can fly in the direction $\omega_k(t)$ at a speed $v_k(t)$, such that its coordinates are updated to $x_k(t + 1) = x_k(t) + v_k(t)\Delta t\cos(\omega_k(t))$ and $y_k(t + 1) = y_k(t) + v_k(t)\Delta t\sin(\omega_k(t))$. To reflect real-world constraints, the speed and direction of UAV $k$ are bounded, i.e., $v_k(t) \leq v_{\text{max}}$ and $\omega_k(t) \in [0, 2\pi)$. To avoid collision among different UAVs, the distance between UAV $k$ and UAV $k'$ should be no less than a protection distance $d_{c}$, i.e.,
\begin{equation}
\label{Eq:collision_limitation}
d_{k,k'}(t) \geq d_{c}, k, k' \in \mathcal{K}, k \neq k', t \in \mathcal{T},
\end{equation}
where $d_{k,k'}(t) = \Vert u_k(t) - u_{k'}(t) \Vert_2$ denotes the Euclidean distance between UAV $k$ and UAV $k'$.

The total energy consumption of UAVs during operation consists primarily of communication and propulsion components, with the latter being dominant\cite{li2023multi}. Accordingly, this study focuses on propulsion energy consumption and neglects the relatively minor communication-related costs. We consider a set of rotary-wing UAVs, and when UAV $k$ flies at a speed of $v_k(t)$ within a two-dimensional (2D) horizontal plane, its propulsion power consumption is given by \cite{zeng2019energy}
\begin{align}
	P_k(v_k(t))& = \frac{1}{2}d_{0}\rho_a s_{\text{sol}} A v_k(t)^3+ P_0\left(1+\frac{3v_k(t)^2}{v^2_{\text{tip}}}\right) \nonumber \\
	&+P_1\left(\sqrt{1+\frac{v_k(t)^4}{4v_{0}^4}}-\frac{v_k(t)^2}{2v^2_{0}}\right)^{\frac{1}{2}},
\end{align}
where $d_0$, $\rho_a$, $s_{\text{sol}}$ and $A$ denote the fuselage drag ratio, air density, rotor solidity and rotor disc area, respectively. $P_0$ and $P_1$ denote the power associated with the blade profile and induced power during hovering, respectively. $v_0$ represents the average rotor-induced velocity during hovering and $v_{\text{tip}}$ is the tip speed of the rotor blade.

Based on the energy consumption model of a rotary-wing UAV flying in a 2D plane derived in the work~\cite{yang2019energy}, the approximate model of propulsion energy consumption in the time slot $t$ is modeled as
\begin{equation}
	\label{Eq:Energy consumption}
	E_k(t) \approx \sum_{t \in \mathcal{T}} P_k(t) \Delta t.
\end{equation}

\subsection{Channel Model}
We introduce the Air-to-Ground (A2G) channel models to capture the communication dynamics within the HetUAVNs. In practice, acquiring the CSI of Eves is challenging, as they are typically passive and non-cooperative. One feasible method is to infer the CSI of the Eve by using the local oscillator signal leaked from the radio frequency front end \cite{zhao2025joint}, \cite{mukherjee2012detecting}, \cite{peng2022deep}. Therefore, for the sake of tractability, this paper focuses on the ideal CSI scenario, where perfect eavesdropper CSI is assumed to be available at the UAVs. The complex-valued channel coefficients between the UAV and GT/Eve are denoted by $\mathbf{h}_{k,x}\in \mathbb{C}^{M \times 1}$, which includes both large-scale fading and small-scale fading. To account for more practical considerations, the large-scale fading of A2G channels are modeled as a combination of LoS and non-LoS (NLoS) components.

Specifically, let $P^{\text{LoS}}_{k,x}(t)$ denote the probability that the channel between UAV $k$ and GT/Eve $x$ is the LoS channel in the time slot $t$, where $x \in \{\mathcal{I}, \mathcal{E}\}$. The probability of an NLoS channel is then $P^{\text{NLoS}}_{k,x}(t)=1-P^{\text{LoS}}_{k,x}(t)$. Following the model \cite{al2014optimal}, the LoS probability can be expressed as
$P_{k, x}^{\text{LoS}}(t) = \frac{1}{1 + a \exp(-b[\arcsin(H_{\text{UAV}}/d_{k, x}(t)) - a])}$,
where $a$ and $b$ are the S-curve parameters related to the actual propagation environment. Consequently, the path loss between the UAV $k$ and GT/Eve $x$ can be expressed as $\ell_{k, x}(t)\! = \!P_{k, x}^{\text{LoS}}(t)\times \eta^{\text{LoS}} \! +  P_{k, x}^{\text{NLoS}}(t)\times \eta^{\text{NLoS}} + \text{FL}_{k,x}(t)$,
where $\eta^{\text{LoS}}$ and $\eta^{\text{NLoS}}$ represent the average additional path loss of the LoS link and the NLoS link, respectively. Additionally, $\text{FL}_{k,x}(t) = 20\log_{10}(4\pi f_c d_{k,x}(t)/c)$ is the free space path loss, with $f_c$ being the carrier frequency and $c$ the speed of light.

On the other hand, the small-scale fading from UAV $k$ to GT/Eve $x$, denoted by $\hat{\mathbf{h}}_{k,x}(t) \in \mathbb{C}^{M \times 1}$, is modeled to follow an i.i.d. Rayleigh distribution. Hence, the A2G channel between UAV $k$ and GT/Eve $x$ can be modeled as $\mathbf{h}_{k,x}(t)=\sqrt{10^{-\frac{1}{10} \times \ell_{k, x}(t)}} \, \hat{\mathbf{h}}_{k,x}(t)$.

\subsection{Heterogeneous Service Models}
In HetUAVNs, UAVs have different payloads and computing capabilities, which results in each UAV $k \in \mathcal{K}$ having different coverage ranges $C_{k}^{r}$ and service capacities $N^s_k$. To characterize the coverage relationships between UAVs and GTs or Eves in each time slot $t$, we define a binary coverage matrix $\mathbf{A}^{\Delta}(t) \in \{0,1\}^{N_{\mathcal{K}} \times N_{\Delta}}$, modeled as $\mathbf{A}^{\Delta}_{k,x}(t) = \mathbf{1}[d_{k,x}(t) \leq C_k^r]$, where $\Delta = \mathcal{I}$ if $x \in \mathcal{I}$, and otherwise $\Delta = \mathcal{E}$.

Because the service capacity of UAVs is limited, each UAV only establishes a communication connection with the GTs with better channel quality within the coverage range. We formalize this relationship using a scheduling matrix $\mathbf{S}^{\mathcal{I}}(t) \in \{0,1\}^{N_{\mathcal{K}} \times N_{\mathcal{I}}}$, where $\mathbf{S}^{\mathcal{I}}_{k,i}(t) = 1$ if GT $i$ is assigned to UAV $k$ in the time slot $t$ and $0$ otherwise. Accordingly, this scheduling needs to satisfy the service capacities constraint
\begin{equation}
	\label{Eq:service capacity limit}
	\sum\limits_{i \in \mathcal{I}} \mathbf{S}_{k,i}^{\mathcal{I}}(t) \leq N^s_k, \forall k \in \mathcal{K}, t \in \mathcal{T}.
\end{equation}
In addition, each GT can be scheduled to at most one UAV, i.e.,
\begin{equation}
	\label{Eq:GT limit}
	\sum\limits_{k\in \mathcal{K}} \mathbf{S}_{k,i}^{\mathcal{I}}(t) \leq 1, \forall i \in \mathcal{I}, t \in \mathcal{T}.
\end{equation}

\subsection{Transmission Model}

Recently, RSMA, built upon the concept of rate-splitting (RS), has been recognized as a promising physical layer transmission paradigm for non-orthogonal transmission, interference management and multiple access strategies in 6G\cite{mao2022rate}. Therefore, we introduce RSMA into HetUAVNs to fully utilize its potential in complex interference management and resource allocation, thereby improving the communication performance of the entire system.
All derivations in the section are performed within a single time slot, with the time symbol $t$ omitted for simplicity.

According to the RS principle, the message $\mathcal{W}_{k,i}$ intended to GT $i$ from UAV $k$ is split into a common part $\mathcal{W}^c_{k,i}$ and a private part $\mathcal{W}_{k,i}^p$, where $i \in \mathcal{I}_k$, and $\mathcal{I}_k$ denotes the set of GTs assigned to UAV $k$. The common parts of GTs in $\mathcal{I}_k$ are encoded together into a common stream $\mathbf{s}^c_k$ using a shared codebook\cite{yang2021optimization}, while each private part $\mathcal{W}_{k,i}^p$ is individually encoded into its corresponding private stream $\mathbf{s}^p_{k,i}$. After the stream $\mathbf{s}_k=[s_k^c, s_{k, 1}^p, \ldots, s_{k, |\mathcal{I}_k|}^p]^T$ are precoded using $\mathbf{P}_k=[\mathbf{p}_k^c, \mathbf{p}_{k,1}^p, \ldots, \mathbf{p}_{k, |\mathcal{I}_k|}^p] \in \mathbb{C}^{M \times (|\mathcal{I}_k| + 1)}$ at the antennas, the signal $\mathbf{x}_k$ transmitted by UAV $k$ is given by $\mathbf{x}_k = \mathbf{P}_k \mathbf{s}_k = \mathbf{p}_k^c s_k^c + \sum\nolimits_{i \in \mathcal{I}_k} \mathbf{p}_{k,i}^p s^p_{k,i}, $ where $\mathbf{p}_k^c$ and $\mathbf{p}_{k,i}^p$ are the precoding vector for the common stream and the private stream, respectively. Supposing that $\mathbb{E}[\mathbf{s}_k\mathbf{s}^H_k]=\mathbf{I}$, we have $\text{Tr}(\mathbf{P}_k\mathbf{P}_k^H) \leq P_{\max}$ and $P_{\max}$ is the transmit power constraint at transmit UAV $k$. Accordingly, the received signal at GT/Eve $x$ from UAV $k$ is
\begin{equation}
	\label{Eq: recv_signal}
	y_{k,x} = \mathbf{h}_{k,x}^H\mathbf{x}_k + \sum\limits_{k'\in \mathcal{K}\backslash \{k\}}\mathbf{A}^{\Delta}_{k',x} \mathbf{h}^H_{k',x} \mathbf{x}_{k'}  + n_{x},
\end{equation}
where the second term on the RHS of~\eqref{Eq: recv_signal} is the inter-system interference when the node $x$ lies in the coverage range of different UAVs, and $n_{x} \sim \mathcal{CN}(0, \sigma^2_{x})$ represents the AWGN at node $x$.

Upon receiving the signal, each GT first decodes the common stream $s^c_k$  to retrieve the associated common message $\mathcal{W}^c_{k, i}$ by treating all private streams as noise. Hence, the corresponding signal-to-interference-plus-noise ratio (SINR) of GT-$i$ when decoding the common stream $s_k^c$ is given by
\begin{equation}
	\label{Eq:SINR of GT's common part}
	\gamma_{i}^c=\frac{\mathbf{S}^{\mathcal{I}}_{k, i}\left|\mathbf{h}_{k,i}^H\mathbf{p}_k^c\right|^2}
	{\mathbf{S}^{\mathcal{I}}_{k, i}\sum\limits_{i'\in \mathcal{I}_k}\left|\mathbf{h}_{k,i}^H\mathbf{p}^p_{k,i'}\right|^2 + I_i^{\text{in}} + \sigma_i^2},
\end{equation}
where $I_{i}^{\text{in}} =\sum\nolimits_{k' \in \mathcal{K} \backslash \{k\}}\mathbf{A}^{\mathcal{I}}_{k', i} \left| \mathbf{h}^H_{k',i} \mathbf{P}_{k'} \right|^2$ is the inter-system interference that GT $i$ experiences.

After removing the common part, each GT proceeds to decode its private streams via successive interference cancellation (SIC) \cite{joudeh2016robust}. The corresponding SINR at GT $i$ when decoding its private stream $s^p_{k,i}$ is given by
\begin{equation}
	\label{Eq:SINR of GT's private part}
	\gamma_{i}^p=\frac{\mathbf{S}^{\mathcal{I}}_{k,i}\left|\mathbf{h}_{k,i}^H\mathbf{p}^p_{k,i}\right|^2}
	{\mathbf{S}^{\mathcal{I}}_{k,i}\sum\limits_{i' \in \mathcal{I}_k \backslash \{i\}} \left|\mathbf{h}_{k,i}^H\mathbf{p}^p_{k,i'}\right|^2 + I_{i}^{\text{in}} + \sigma_i^2},
\end{equation}

Similarly, the SINR at Eve $e$ when attempting to decode the common stream $s^c_k$ from UAV $k$ is given by
\begin{equation}
	\label{Eq:SINR of Eve's common part}
	\gamma_{e,i}^c=\frac{\mathbf{A}^{\mathcal{E}}_{k,e}\left|\mathbf{h}_{k,e}^H\mathbf{p}_k^c\right|^2}
	{\mathbf{A}^{\mathcal{E}}_{k,e}\sum\limits_{i' \in \mathcal{I}_k}\left|\mathbf{h}_{k,e}^H\mathbf{p}^p_{k,i'}\right|^2 + I_{e,i}^{\text{in}} + \sigma_e^2}, i \in \mathcal{I}_k,
\end{equation}
where $I^{\text{in}}_{e,i}=\sum\nolimits_{k' \in \mathcal{K} \backslash \{k\}}\mathbf{A}_{k',e}^{\mathcal{E}} \left|\mathbf{h}^H_{k',e}\mathbf{P}_{k'}\right|^2$ represents the inter-system interference caused by other UAVs to Eve $e$. To reduce the likelihood of private streams being decoded by Eve, the rate of the common stream from UAV to GT is designed to be higher than the achievable rate for Eve.  Then, the SINR of Eve $e$ when attempting to decode the private stream $\mathbf{s}^p_{k,i}$ of GT $i$ from UAV $k$ is given by
\begin{equation}
	\label{Eq:SINR of Eve's private part}
	\gamma^p_{e,i}\!=\!\frac{\mathbf{A}_{k,e}^{\mathcal{E}}\left|\mathbf{h}_{k,e}^H\mathbf{p}^p_{k,i}\right|^2}
	{\mathbf{A}_{k,e}^{\mathcal{E}}\left(\left|\mathbf{h}_{k,e}^H \mathbf{p}_k^c\right|^2 \!\!\! +\!\!\! \sum\limits_{i'\in \mathcal{I}_k \backslash \{i\}}\left|\mathbf{h}^H_{k,e}\mathbf{p}^p_{k,i'}\right|^2\right) + I_{e,i}^{\text{in}} + \sigma_e^2}.
\end{equation}

\subsection{Multi-Objective Problem Formulation}
We formulate a MOO framework that jointly optimizes UAV trajectories and transmission strategies to trade off secrecy and energy efficiency, with the two objectives defined as follows.

\subsubsection{\textbf{Optimization Objective $1$ (Secrecy Rate Maximization)}}
Based on the SINRs in~\eqref{Eq:SINR of GT's common part} and~\eqref{Eq:SINR of GT's private part}, the achievable rates of common and private messages at GT $i$  are $R^c_{i}(t) = \log_2(1 + \gamma_{i}^c(t))$ and $R^p_{i}(t) = \log_2(1 + \gamma_{i}^p(t))$, respectively. Similarly, the achievable rates of common and private messages at Eve $e$ are $R^c_{e,i}(t) = \log_2(1 + \gamma_{e,i}^c(t))$ and $R^p_{e, i}(t) = \log_2(1 + \gamma_{e, i}^p(t))$, where $\gamma_{e, i}^c(t)$ and $\gamma_{e, i}^p(t)$ are defined in~\eqref{Eq:SINR of Eve's common part} and~\eqref{Eq:SINR of Eve's private part}.

According to the standard RSMA mechanism \cite{mao2022rate}, the common secrecy rate for all GTs served by UAV $k$ at time $t$ is given by $R^{\text{sr,c}}_k(t) = \left[ R_{i^*}^c(t) - R_{e^*, i}^c(t) \right]^+$, where $i^* = \arg \min_{i \in \mathcal{I}_k} R^c_i(t)$ denotes the GT with the lowest achievable common rate among those served by UAV $k$, and $e^* = \arg \max_{e \in \mathcal{E}_k} R^c_{e, i}(t)$ represents the Eve that achieves the highest eavesdropped common rate from UAV $k$. Similarly, the private secrecy rate for GT $i$ is expressed as $R^{\text{sr,p}}_{k,i}(t) = \left[ R_i^p(t) - R_{e^*,i}^p(t) \right]^+$, where $e^* = \arg \max_{e\in\mathcal{E}_k} R_{e,i}^p(t)$  is the Eve that maximizes the eavesdropped private rate.
To better evaluate the overall secrecy performance of the system, we model the problem as maximizing the worst-case secrecy rate among all GTs according to \cite{nasir2016secrecy}. Thus, the objective $1$ is formulated as
\begin{align}
	\label{Eq:Objective_1}
	f_1(\boldsymbol{\omega}, \mathbf{v}, \mathbf{P}) \triangleq \min_{k \in \mathcal{K}, i \in \mathcal{I}_k} \left( \alpha_{k,i} R_{k}^{\text{sr,c}}(t) + R_{k,i}^{\text{sr,p}}(t) \right)
\end{align}
where $\boldsymbol{\omega} \triangleq \{\omega_k(t) | k \in \mathcal{K}, t \in \mathcal{T}\}$, $\boldsymbol{v} \triangleq \{v_k(t)|k\in\mathcal{K}, t\in\mathcal{T}\}$ and $\mathbf{P} \triangleq \{\mathbf{P}_k(t) | k \in \mathcal{K}, t \in \mathcal{T}\}$ are the flight direction, speed and precoding matrices of UAVs, respectively. And $\mathcal{E}_k$ represents the set of Eves that eavesdrop on UAV $k$.

The weighting factor $\alpha_{k, i} \in [0, 1]$ is a constant that governs the allocation of the common secrecy rate among the GTs associated with UAV $k$, satisfying $\sum\nolimits_{i \in \mathcal{I}_k} \alpha_{k, i} = 1$ \cite{fu2020robust}. For simplicity and fairness, the common secrecy rate is thus divided equally among all associated GTs by setting $\alpha_{k, i} = 1 / |\mathcal{I}_k|$.

\subsubsection{\textbf{Optimization Objective $2$ (Propulsion Energy Consumption Minimization)}}
Based on the propulsion energy model in~\eqref{Eq:Energy consumption}, the second optimization objective is formulated as minimizing the total propulsion energy consumption of all UAVs over the entire time horizon of $N_{\mathcal{T}}$ time slots. This objective can be expressed as
\begin{equation}
    \label{Eq:Objective_2}
	f_2(\boldsymbol{\omega}, \boldsymbol{v}, \mathbf{P}) \triangleq  \sum_{k \in \mathcal{K}} E_k(N_{\mathcal{T}}).
\end{equation}

Based on the two optimization objectives presented in~\eqref{Eq:Objective_1} and~\eqref{Eq:Objective_2}, the MOO problem in secure HetUAVNs is formulated as follows \cite{zhang2024multi}, \cite{song2022evolutionary}:
\begin{subequations}
	\label{Eq:MOP1}
	\begin{align}
		\textbf{P1:} \quad \max_{\boldsymbol{\omega}, \boldsymbol{v}, \mathbf{P}} & \quad F  \triangleq \{f_1, -f_2\}, \label{Eq:MOP1_Objective}\\
		\text{s.t.} \quad & u_k(t) \in [0, D]^2, \quad \forall k \in \mathcal{K}, t \in \mathcal{T}, \label{Eq:MOP1_C1}\\
		& \omega_k(t) \in [0, 2\pi), \quad\forall k \in \mathcal{K}, t \in \mathcal{T},  \label{Eq:MOP1_C2}\\
		&  v_k(t) \leq v_{\max},  \quad\forall k \in \mathcal{K}, t \in \mathcal{T},  \label{Eq:MOP1_C3}\\
		&  \text{Tr}(\mathbf{P}_k(t)\mathbf{P}_k^H(t)) \leq P_{\max}, \forall k \in \mathcal{K}, t \in \mathcal{T}, \label{Eq:MOP1_C5} \\
		& R_i^c(t) \geq R^c_{e,i}(t), \forall k \in \mathcal{K}, e \in \mathcal{E}, t \in \mathcal{T},  \label{Eq:MOP1_C6}\\
		& \eqref{Eq:collision_limitation},~\eqref{Eq:service capacity limit},~\eqref{Eq:GT limit},
 	\end{align}
\end{subequations}
where~\eqref{Eq:MOP1_C1} ensures that all UAVs remain within the service area for all time slots.~\eqref{Eq:MOP1_C2} regulates the flight direction selection of the UAV.~\eqref{Eq:MOP1_C3} restricts each UAV's speed to be below the maximum speed.~\eqref{Eq:MOP1_C5} imposes a limit on the maximum transmit power of each UAV.~\eqref{Eq:MOP1_C6} reduces the likelihood of Eve decoding private messages.

To solve this deeply coupled and non-convex problem, we propose a hierarchical optimization framework that decomposes the joint optimization of secrecy precoding and heterogeneous UAV trajectories into two tractable sub-problems, deliberately leveraging the distinct strengths of two computational paradigms. The inner layer is a mathematically rigorous non-convex problem, solved under fixed UAV positions by the S2DC algorithm acting as a precise operator embedded in the environment dynamics. The outer layer, in contrast, handles long-term,  globally-aware decision-making under uncertainty, for which we propose an LLM-guided heuristic MARL method to  optimize UAV trajectories. This hybrid of optimization-theoretic and learning-based decision-making has proven effective for complex coupled problems in wireless networks~\cite{hassan2024spaceris},~\cite{zhao2024joint},~\cite{gao2023dynamic}.

\section{The proposed S2DC for Secrecy Precoding}
In this section, we propose the S2DC algorithm to address the secrecy precoding optimization problem when all UAVs are fixed. The problem is accordingly formulated as
\begin{subequations}
	\label{Eq:MOP2-Precoding-Solve}
	\begin{align}
		\textbf{P2:} \quad \max_{\mathbf{P}}  & \quad F_1 \triangleq \min_{k \in \mathcal{K}, i \in \mathcal{I}_k} \left( \alpha_{k,i} R_{k}^{\text{sr,c}}(t) + R_{k,i}^{\text{sr,p}}(t) \right), \label{Eq:MOP2-Precoding-Solve-Obj} \\
		\text{s.t.} \quad &  \eqref{Eq:MOP1_C5},~\eqref{Eq:MOP1_C6}.
	\end{align}
\end{subequations}

By applying SDR to denote the outer products $\mathbf{P}^c_k \triangleq \mathbf{p}^c_k(\mathbf{p}^c_k)^H$, $\mathbf{P}^p_{k,i} \triangleq \mathbf{p}^p_{k,i}(\mathbf{p}^p_{k,i})^H$, and then $\mathbf{P}^c \triangleq \{\mathbf{P}^c_{k}|k \in \mathcal{K}\}$, $\mathbf{P}^p \triangleq \{\mathbf{P}^p_{k,i}|k \in \mathcal{K}, i \in \mathcal{I}_k\}$, we transform~\eqref{Eq:MOP2-Precoding-Solve-Obj} into
\begin{align}
	\label{Eq:linearization_target}
 \tilde{F}_1(&\mathbf{P}^c, \mathbf{P}^p)  =  \tilde{F}_{1,1}(\mathbf{P}^c, \mathbf{P}^p)  \nonumber \\
 & + \tilde{F}_{1,2}(\mathbf{P}^c, \mathbf{P}^p)- (\tilde{F}_{1,3}(\mathbf{P}^c, \mathbf{P}^p) +  \tilde{F}_{1,4}(\mathbf{P}^c, \mathbf{P}^p)),
\end{align}
where $\tilde{F}_{1,1}$, $\tilde{F}_{1,2}$, $\tilde{F}_{1,3}$, and $\tilde{F}_{1,4}$ are respectively given by
\begin{align}
	\tilde{F}_{1,1} \triangleq & \log_2(\phi^c_{i^*} + \mathbf{h}_{k,i^*}^H \mathbf{P}^c_k \mathbf{h}_{k,i^*}) + \log_2(\phi^c_{e^*,i}),
	\\
	\tilde{F}_{1,2} \triangleq & \log_2(\phi^p_i+\mathbf{h}_{k,i}^H\mathbf{P}^p_{k,i}\mathbf{h}_{k,i}) + \log_2(\phi^p_{e^*,i}),
	\\
	\tilde{F}_{1,3} \triangleq & \log_2(\phi^c_{i^*}) + \log_2(\phi^c_{e^*,i} + \mathbf{h}_{k,e^*}^H\mathbf{P}_k^c\mathbf{h}_{k,e^*}),
	\\
	\tilde{F}_{1,4} \triangleq & \log_2(\phi^p_i) + \log_2(\phi^p_{e^*,i} + \mathbf{h}_{k,e^*}^H\mathbf{P}^p_{k,i}\mathbf{h}_{k,e^*}),
\end{align}
and
\begin{align}
	\phi^c_{i^*}(\mathbf{P}^c, \mathbf{P}^p) &\triangleq \sum\limits_{i'\in \mathcal{I}_k}\mathbf{h}_{k,i^*}^H\mathbf{P}^p_{k,i'}\mathbf{h}_{k,i^*} + \tilde{I}^{\text{in}_{i^*}} +\sigma_{i^*}^2, \\
	\phi^p_i(\mathbf{P}^c, \mathbf{P}^p) &\triangleq \sum\limits_{i' \in \mathcal{I}_k \backslash \{i\}} \mathbf{h}_{k,i}^H\mathbf{P}^p_{k,i'}\mathbf{h}_{k,i} + \tilde{I}^{\text{in}}_i +\sigma_i^2, \\
	\phi^c_{e^*,i}(\mathbf{P}^c, \mathbf{P}^p) &\triangleq \sum\limits_{i' \in \mathcal{I}_k}\mathbf{h}_{k,e^*}^H\mathbf{P}^p_{k,i'}\mathbf{h}_{k,e^*} + \tilde{I}^{\text{in}}_{e^*,i} + \sigma_{e^*}^2, \\
	\phi^p_{e^*,i}(\mathbf{P}^c, \mathbf{P}^p) &\triangleq \mathbf{h}_{k,e^*}^H \mathbf{P}_k^c\mathbf{h}_{k,e^*} + \sum\limits_{i'\in \mathcal{I}_k \backslash \{i\}}
	\mathbf{h}^H_{k,e^*}\mathbf{P}^p_{k,i'}\mathbf{h}_{k,e^*} \nonumber \\
	&\quad + \tilde{I}^{\text{in}}_{e^*,i} + \sigma_{e^*}^2.
\end{align}

Since $\tilde{F}_{1,1} \sim \tilde{F}_{1,4}$ are convex in $(\mathbf{P}^c, \mathbf{P}^p)$, $\tilde{F}_1$ is a D.C. function, and problem~\eqref{Eq:MOP2-Precoding-Solve} can be equivalently transformed into
\begin{subequations}
	\label{Eq:MOP2-Precoding-Solve-SDR}
	\begin{align}
		\max_{\mathbf{P}^c, \mathbf{P}^p}  & \quad \tilde{F}_{1}, \label{Eq:MOP2_Objective} \\
		\text{s.t.} \quad &  \text{Tr}(\mathbf{P}^c_k) + \sum\nolimits_{i \in \mathcal{I}_k} \text{Tr}(\mathbf{P}^p_{k,i}) \leq P_{\text{max}}, \label{Eq:MOP2_C1} \\
		& \tilde{F}_{1,1} - \tilde{F}_{1,3} \geq 0, \label{Eq:MOP2_C2} \\
		& \mathbf{P}^c_k \succeq 0, \quad \mathbf{P}^p_{k,i} \succeq 0,  \label{Eq:MOP2_C3} \\
		& \text{rank}(\mathbf{P}^c_k)=1, \quad \text{rank}(\mathbf{P}^p_{k,i})=1, \label{Eq:MOP2_C4} \\
		& \forall k \in \mathcal{K}, i \in \mathcal{I}_k, e \in \mathcal{E}_k, \label{Eq:MOP2_C5}
	\end{align}
\end{subequations}
where constraints~\eqref{Eq:MOP2_C1} and~\eqref{Eq:MOP2_C3} are convex functions, while the~\eqref{Eq:MOP2_Objective} and~\eqref{Eq:MOP2_C2} are D.C. functions. By dropping the rank-one nonconvex constraints~\eqref{Eq:MOP2_C4}, the problem~\eqref{Eq:MOP2-Precoding-Solve-SDR} can be solved directly via D.C. iterations \cite{kha2011fast}.

{\footnotesize
	\begin{algorithm}
		\caption{Maximizing the secrecy rate using SDR and D.C. iterations (S2DC).}
		\label{Alg:iterative-algo}
		\DontPrintSemicolon
		\SetAlgoLined
		\KwIn{Channel matrices.}
		\KwOut{Optimized precoding $\mathbf{P}^*$.}
		
		\textbf{Initialization}: Set the maximum numbers of iterations $N_{\text{iter}}$, the penalty parameter $\mu$, the iteration index $l=1$, a feasible point $\mathbf{P}^0$ and the maximum tolerance $\epsilon$;
		
		Transform the problem~\eqref{Eq:MOP2-Precoding-Solve} into a semidefinite programming (SDP);
		
		Transform the non-convex rank-one constraint~\eqref{Eq:MOP2_C4} into~\eqref{Eq: rank-one constrain equivalent transformation 1} and~\eqref{Eq: rank-one constrain equivalent transformation 2};
		
		By accurately penalizing non-convex constraints, the problem is transformed into~\eqref{Eq:MOP4};

		\Repeat{$|\mathcal{F}^{(l)} - \mathcal{F}^{(l - 1)}| \leq \epsilon$ \text{or} $l \geq N_{\text{iter}}$.}{
				For each UAV $k$, determine the indices $i^*$ and $e^*$ using the current solution $\mathbf{P}^{(l - 1)}$;
			
				Given the fixed indices $i^*$ and $e^*$, solve~\eqref{Eq:MOP-Precoding-Solve-dc-iter-cvx} to obtain $\mathbf{P}^{l}$ by exploiting the CVX toolbox;
			
			Set $l := l + 1$;
		}
	\end{algorithm}
}

However, the rank-one constraint~\eqref{Eq:MOP2_C4} is non-convex. According to \cite{nasir2016secrecy}, this constraint can be equivalently written as
\begin{align}
	\text{Tr}(\mathbf{P}^{c}_k) - \lambda_{\text{max}}(\mathbf{P}^c_k) & \leq 0, k \in \mathcal{K}, \label{Eq: rank-one constrain equivalent transformation 1} \\
	\text{Tr}(\mathbf{P}^p_{k,i})-\lambda_{\text{max}}(\mathbf{P}^p_{k,i}) & \leq 0, k \in \mathcal{K}, i \in \mathcal{I}_k, \label{Eq: rank-one constrain equivalent transformation 2}
\end{align}
where $\lambda_{\text{max}}(\mathbf{P}^c_k)$($\lambda_{\text{max}}(\mathbf{P}^p_{k,i})$, resp.) is the maximal eigenvalue of $\mathbf{P}^{c}_k$($\mathbf{P}^p_{k,i}$, resp.). Using the exact penalty technique in \cite{phan2012nonsmooth}, this non-convex constraint is introduced into the objective function in the form of a penalty term, thus reformulating the problem~\eqref{Eq:MOP2-Precoding-Solve-SDR} as
\begin{subequations}
	\label{Eq:MOP4}
	\begin{align}
		\max_{\mathbf{P}^c, \mathbf{P}^p} \,\, & \min_{k \in \mathcal{K}, i \in \mathcal{I}_k, e \in \mathcal{E}_k} \tilde{F}_{1} + \mu \Big[\sum_{k \in \mathcal{K}}(\lambda_{\text{max}}(\mathbf{P}^c_k)-\text{Tr}(\mathbf{P}^{c}_k)) \nonumber \\
		& +\sum_{k\in\mathcal{K}}\sum_{i\in\mathcal{I}_k}(\lambda_{\text{max}}(\mathbf{P}^p_{k,i})-\text{Tr}(\mathbf{P}^p_{k,i}))\Big], \\
		\text{s.t.} \quad &  \eqref{Eq:MOP2_C1}-\eqref{Eq:MOP2_C3},~\eqref{Eq:MOP2_C5},
	\end{align}
\end{subequations}
for penalty parameter $\mu > 0$, which is again maximization of a D.C. function subject to convex constraints. Therefore, the D.C iteration technique can be used to generate feasible points ($\mathbf{P}^{c,(l + 1)}$, $\mathbf{P}^{p,(l + 1)}$) from the incumbent ($\mathbf{P}^{c,(l)}$, $\mathbf{P}^{p,(l)}$) by solving a convex program 
\begin{subequations}
	\label{Eq:MOP5-Precoding-Solve-dc-iter}
	\begin{align}
		\max_{\mathbf{P}^c, \mathbf{P}^p}  \,\, \mathcal{F} =& \Big\{\min_{k \in \mathcal{K}, i \in \mathcal{I}_k, e \in \mathcal{E}_k} \Big[\tilde{F}_{1,1} + \tilde{F}_{1,2} - (\tilde{F}^{(l)}_{1,3} + \tilde{F}^{(l)}_{1,4})\Big], \nonumber \\
		& + \mu \Big[\sum_{k \in \mathcal{K}}(\lambda^{(l)}_k(\mathbf{P}^c_k)-\text{Tr}(\mathbf{P}^{c}_k)) \nonumber \\
		& +\sum_{k\in\mathcal{K}}\sum_{i\in\mathcal{I}_k}(\lambda^{(l)}_k(\mathbf{P}^p_{k,i})-\text{Tr}(\mathbf{P}^p_{k,i}))\Big] \Big\} \\
		\text{s.t.} \qquad  & \tilde{F}_{1,1} - \tilde{F}_{1,3}^{(l)} \geq 0, \label{Eq:secrecy_con} \\
        & \,\,  \eqref{Eq:MOP2_C1},~\eqref{Eq:MOP2_C3},~\eqref{Eq:MOP2_C5},
	\end{align}
\end{subequations}
where
\begin{align}
		\lambda_{k}^{(l)}(\mathbf{P}^c_k) & \! = \! \lambda_{\max}(\mathbf{P}^{c,(l)}_k) \! + \!(\mathbf{\bar{p}}^{c,(l)}_k)^H(\mathbf{P}^c_k \! - \! \mathbf{P}^{c,(l)}_k)\mathbf{\bar{p}}^{c,(l)}_k, \\
		\lambda_{k,i}^{(l)}(\mathbf{P}^p_{k,i}) & \! = \! \lambda_{\max}(\mathbf{P}^{p,(l)}_{k,i}) \! + \! (\mathbf{\bar{p}}^{p,(l)}_{k,i})^H(\mathbf{P}^p_{k,i}\!\!-\!\!\mathbf{P}^{p,(l)}_{k,i})\mathbf{\bar{p}}^{p,(l)}_{k,i},	
\end{align}
and $\mathbf{\bar{p}}^{c,(l)}_k$ (resp. $\mathbf{\bar{p}}^{p,(l)}_{k,i}$) is the normalized eigenvector corresponding to $\lambda_{\text{max}}(\mathbf{P}^c_k)$ (resp. $\lambda_{\text{max}}(\mathbf{P}^p_{k,i})$). $\tilde{F}^{(l)}_{1,m}(\mathbf{P}^c, \mathbf{P}^p)$ denotes the first-order Taylor expansion of $\tilde{F}_{1,m}(\mathbf{P}^c, \mathbf{P}^p)$ at the $l$-th iteration point ($\mathbf{P}^{c,(l)}$, $\mathbf{P}^{p,(l)}$), defined as
\begin{align}
	& \tilde{F}^{(l)}_{1,m}(\mathbf{P}^c, \mathbf{P}^p) = \tilde{F}_{1,m}(\mathbf{P}^{c,(l)}, \mathbf{P}^{p,(l)}) \nonumber \\
	& + \langle \nabla\tilde{F}_{1,m}(\mathbf{P}^{c,(l)}, \mathbf{P}^{p,(l)}), (\mathbf{P}^{c}, \mathbf{P}^{p}) - (\mathbf{P}^{c,(l)}, \mathbf{P}^{p,(l)}) \rangle,
\end{align}
for $m \in \{3,4\}$. To render the problem tractable, we introduce an epigraph variable and equivalently reformulate problem~\eqref{Eq:MOP5-Precoding-Solve-dc-iter} as
\begin{subequations}
	\label{Eq:MOP-Precoding-Solve-dc-iter-cvx}
	\begin{align}
		\max_{\mathbf{P}^c, \mathbf{P}^p} & \quad \varphi \\
		\text{s.t.} \quad & \mathcal{F}  \geq \varphi,  \\
		 & \eqref{Eq:MOP2_C1},~\eqref{Eq:MOP2_C3},~\eqref{Eq:MOP2_C5},~\eqref{Eq:secrecy_con},
	\end{align}
\end{subequations}
which can be solved efficiently by the MOSEK solver in the CVX toolbox. For the outer product matrix output by the S2DC algorithm, we use the standard rank-one approximation method, that is, we perform eigenvalue decomposition on each matrix $\mathbf{P}^c_k$ or $\mathbf{P}^p_{k,i}$ and take its principal eigenvector to reconstruct the precoding vector\cite{9739676}.

Formally, we summarize the S2DC in Algorithm~\ref{Alg:iterative-algo}. 
The theoretical complexity of solving this problem using MOSEK solver is $\mathcal{O}\left(\sqrt{q} \log(1/\epsilon) (pq^3 + p^2q^2 + p^3)\right)$, where $p$ represents the number of constraints and $q$ denotes the number of decision variables. In this paper, the values of $p$ and $q$ depend primarily on the service capacity of the UAV. Hence, the worst computational complexity of S2DC can be estimated as $\mathcal{O}_{\text{S2DC}}=(\sum N_k^s)^{4.5} \log(1/\epsilon)$ \cite{10718357}.

\begin{figure*}
	\centering
	\includegraphics[width=0.9\linewidth]{"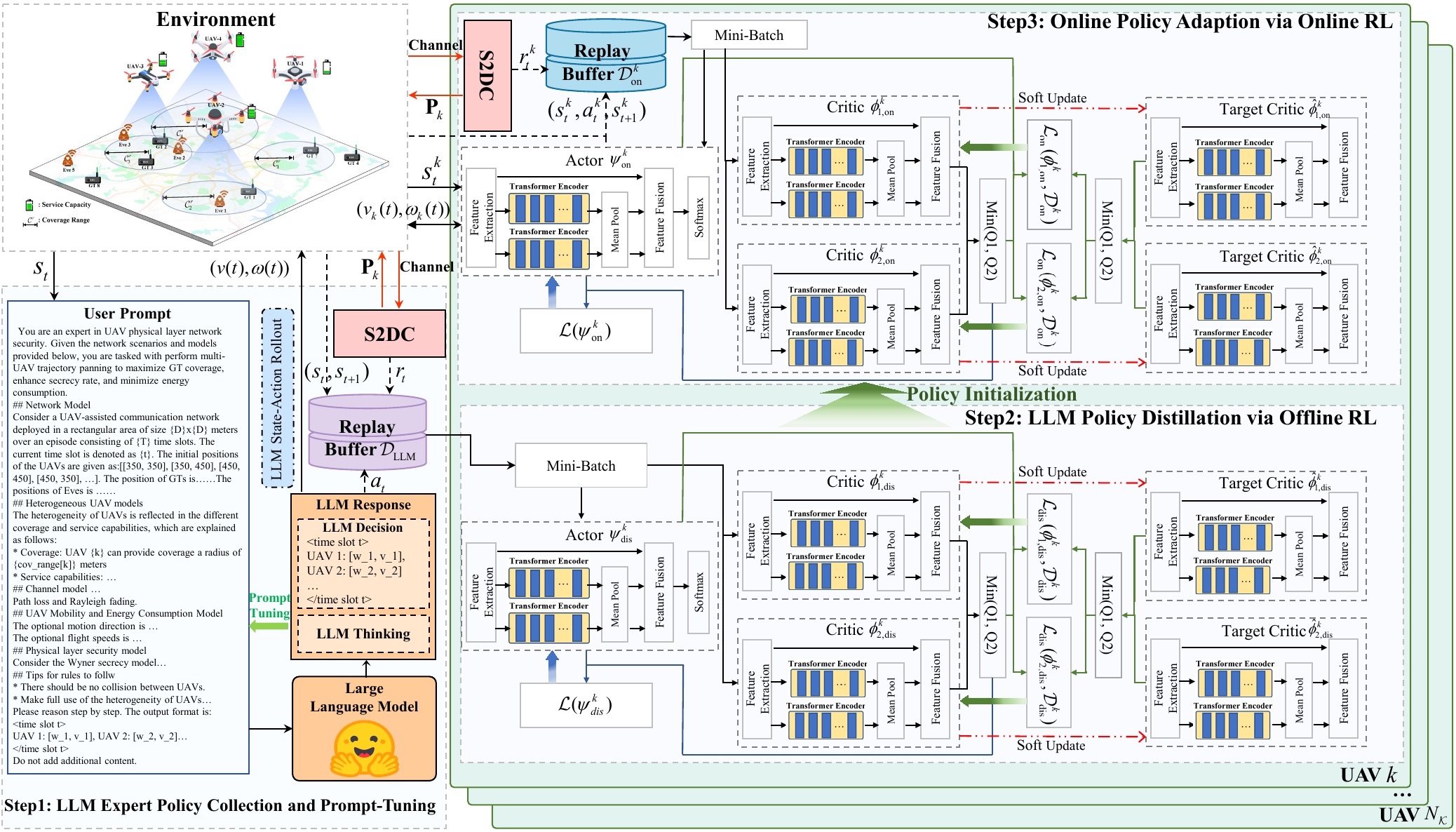"}
	\caption{Framework of the LLM-HeMARL-S2DC algorithm in secure HetUAVNs.}
	\label{Fig:algo_framework}
\end{figure*}

\section{The proposed LLM-HeMARL for Collaborative Trajectories Design}

Based on the problem decomposition presented in the previous section, we investigate the outer-layer collaborative trajectories design, while incorporating the inner-layer S2DC based secrecy precoding. Accordingly, the problem is formulated as 
\begin{subequations}
	\label{Eq:MOP-Trajectory-Solve}
	\begin{align}
		\textbf{P3:} \quad \max_{\boldsymbol{\omega}, \boldsymbol{v}}  & \quad F_2 \triangleq \{f_1, -f_2\} \\
		\text{s.t.} \quad & \mathbf{P}_k=\text{S2DC}(\mathbf{h}_{k,x}), x \in \{\mathcal{I}, \mathcal{E}\}, k \in \mathcal{K}, \label{Eq:secrecy_precoding}\\
		& \eqref{Eq:MOP1_C1}-\eqref{Eq:MOP1_C3},~\eqref{Eq:collision_limitation},~\eqref{Eq:service capacity limit},~\eqref{Eq:GT limit},~\label{Eq:SR_C7}
	\end{align}
\end{subequations}
where~\eqref{Eq:secrecy_precoding} represents the secrecy precoding obtained via the S2DC, which is computed based on the channel conditions under fixed UAV positions. To solve this problem, we propose an LLM-driven heuristic MARL (LLM-HeMARL) to solve~\eqref{Eq:MOP-Trajectory-Solve}, an adaptive policy infusion and distillation framework that incorporates LLM expert policy into MARL for collaborative UAV trajectories design. As shown in Fig.~\ref{Fig:algo_framework}, the proposed algorithm comprises three stages: LLM expert policy collection and prompt fine-tuning, LLM Policy distillation via offline RL, and online policy adaptation via online RL. Next, we first formulate problem~\eqref{Eq:MOP-Trajectory-Solve} as a Markov decision process (MDP), and then describe the three steps of the algorithm in detail.

\subsection{MDP Formulation}

The problem is formulated as an MDP defined by the tuple $(\mathcal{S}, \mathcal{A}, \mathcal{P}, \mathcal{R}, \gamma)$, where $\mathcal{S}$, $\mathcal{A}$, $\mathcal{P}$, $\mathcal{R}$, and $\gamma \in [0,1]$ denote the state space, action space, transition probability, reward function, and discount factor, respectively. The state, action, and reward are detailed below.

\subsubsection{\textbf{State Space} $\mathcal{S}$}
The state space is designed to capture the key spatial and environmental factors that influence system performance. Specifically, the coordinates of the UAVs, GTs and Eves are contained, as they directly determine the channel conditions. And the UAV can directly obtain this information through synthetic aperture radar\cite{li2021joint}, reducing the communication overhead of obtaining other features. To better characterize spatial relationships, in this work, we adopt relative positions to represent all positional relations in the system. As such, the state $s^k_t$ of UAV $k$ in the time slot $t$ can be described as 
$s^k_t = \langle \!\{u_k(t)\! - \!u_l(t)\}_{l \in \mathcal{K} \backslash \{k\}}, \{u_k(t) \!- \!u_i\}_{i \in \mathcal{I}},  \{u_k(t) \!-\! u_e\}_{e\in \mathcal{E}}  \!\rangle$.

\subsubsection{\textbf{Action Space} $\mathcal{A}$}
After obtaining the corresponding state information, each UAV agent selects its action $a^k_t$ following their policy distribution, which can be defined as
$a^k_t = \{v_k(t), \omega_k(t)\}$, where $v_k(t)$ is quantified based on the logarithmic normalization method mentioned in \cite{meng2020power} and is quantized to $v_k(t) \in \{0, \{V_{\text{min}}(\frac{V_{\text{max}}}{V_{\text{min}}})^{\frac{l}{|L| - 2}}|l=0,\ldots, |L| - 2\}\}$, where $|L|$ is the number of selectable velocities. The direction of movement $\omega_k(t) = \{\text{upward}, \text{downward}, \text{left}, \text{right}, \text{still}\}$.

\subsubsection{\textbf{Reward Function} $\mathcal{R}$}
After establishing the states and action spaces, the next step involves defining a reward function $r(s_t, a_t)$ that aligns with the optimization problem’s objectives while satisfying the relevant constraints. To capture the two primary optimization objectives—secrecy rate maximization and propulsion energy consumption minimization—we define two corresponding reward components. The secrecy rate-based reward is given by $r^{\text{sr}}_t = \sum\nolimits_{k \in \mathcal{K}} \sum\nolimits_{i \in \mathcal{I}_k} R_i^{\text{sr}}(t)$, where $R_i^{\text{sr}}(t) = R_i(t) - \max\{R_{e,i}(t)|e\in\mathcal{E}\}$ is the worst-case secrecy rate of GT $i$. Accordingly, the energy consumption-based reward is defined as $r^{\text{ec}}_t = -\sum_{k\in\mathcal{K}}E_k(t)$.

For effective collaboration among UAV agents, all agents share global utility. Thus, the reward function for each UAV is formulated as $r^k(s^k_t, a^k_t) = (w^{\text{sr}} r^{\text{sr}}_t + w^{\text{ec}} r^{\text{ec}}_t) \times \eta^{\text{loc}}_{k,t} - \eta^{\text{col}}_{k,t} \times p^{\text{col}}$,
where $w^{\text{sr}}$ and $w^{\text{ec}}$ denote the weight factors for the two objectives, respectively.
To improve training stability, the weights were selected following the methodologies mentioned in~\cite{zhang2024multi},~\cite{li2024collaborative}, based on the value ranges of their respective reward components, ensuring that all items remain on the same order of magnitude. Additionally, binary indicators $\eta^{\text{loc}}_{k,t},\eta^{\text{col}}_{k,t} \in \{0, 1\}$ are introduced to penalize violations of the flight boundary and collision avoidance constraints, respectively. Here, $p^{\text{col}}$ represents a constant penalty imposed for potential collision risks.

\subsection{LLM Expert Policy Collection and Prompt-Tuning}
The main purpose of this step is to collect expert policy from the LLM by deploying it as an agent that interacts with the environment in a closed loop. Specifically, our framework begins with the manual construction of a comprehensive textual prompt that follows established prompt engineering principles\cite{white2023prompt}. This prompt encapsulates the entire task description, including the initial system configuration, mission objective, channel model, secrecy constraints, operational rules, and any other relevant limitations. Upon receiving the prompt, the LLM performs multi-step reasoning based on the provided initial environmental state and employs a chain-of-thought mechanism\cite{wei2022chain} to generate a detailed internal thought process that leads to a final decision or action. During this phase, we record the initial environmental configuration, the complete reasoning process, and the resulting action selected.

Next, the positions of the UAVs are updated based on LLM's decisions, and the corresponding CSI is obtained. This CSI is then fed into the S2DC module to compute the secrecy precoding. Subsequently, the reward $r_t$ is calculated based on the UAVs' propulsion consumption and secrecy rate.
Meanwhile, we tuned the prompts by analyzing the LLM's reasoning and decision outcomes to reduce hallucinations and improve the reliability of the answers. Finally, using regular expressions, we parse the stored environmental parameters and LLM-generated policies into RL trajectory format, thereby constructing an LLM policy dataset $\mathcal{D}_{\text{LLM}}$, which is formally defined as $\mathcal{D}_{\text{LLM}} = \{(s_t, a_t, r_t, s_{t+1})|a_t \sim \pi_{\text{LLM}}(a_t|s_t)\}$, where $s_t$, $a_t$, $r_t$, and $s_{t+1}$ denote the state, action, reward, and next state at time step $t$, respectively, and $\pi_{\text{LLM}}$ represents the policy implicitly induced by the LLM through its prompting mechanism.

\subsection{LLM Policy Distillation and Online Policy Adaptation}

To obtain end-to-end control policies tailored to the UAV communication environment, we employ offline RL  method to distill the LLM expert policy stored in $\mathcal{D}_{\text{LLM}}$ into a fast policy. Subsequently, the agents equipped with the distilled policy interact with the environment for parameter fine-tuning, thereby adapting  to environmental states not covered in $\mathcal{D}_{\text{LLM}}$. We build on Soft Actor-Critic (SAC)~\cite{haarnoja2018soft} for its balance of exploration and exploitation and its robustness to value overestimation, and extend it to a decentralized multi-agent setting as the Independent Soft Actor-Critic (ISAC) algorithm, in which each agent keeps its own replay buffer to avoid mixing heterogeneous UAV experiences.

\subsubsection{Independent Soft Actor-Critic Algorithm}
Each agent independently maintains its own actor network, critic networks, target critic networks, and experience replay buffer. First, the actor network: parameterized by $\psi$, this network approximates the policy $\pi_{\psi}(a_t | s_t)$, which maps a given state $s_t$ to a distribution over discrete actions. The policy is formally defined as $ \pi_{\psi}(a_t | s_t) = \text{Softmax}(f_{\psi}(s_t)) = \frac{\exp(f_{\psi}(s_t)_{a_t})}{\sum_{a_t^{'} \in \mathcal{A}}\exp(f_{\psi}(s_t)_{a_t^{'}})}$, where $f_{\psi}(s_t)$ denotes the raw output logits from the policy network for state $s_t$. Second, the critic networks: two Q-value approximators $Q_{\phi_1}$ and $Q_{\phi_2}$ with parameters $\phi_1$ and $\phi_2$. Correspondingly, two target critic networks with parameters $\hat{\phi}_1$ and $\hat{\phi}_2$ compute the target Q-values $Q_{\hat{\phi}_1}(s_t,a_t)$ and $Q_{\hat{\phi}_2}(s_t,a_t)$. The dual critic architecture mitigates Q-value overestimation. Third, entropy regularization: a temperature-adjusted entropy term is incorporated into the policy objective to promote exploration during online learning, expressed as $\mathcal{H}(\pi(\cdot | s_t)) = -\sum\nolimits_{a_t \in \mathcal{A}} \pi(a_t | s_t) \log \pi (a_t | s_t)$, 
which encourages diverse action selection to facilitate reward maximization.

According to the components incorporated within the ISAC learning architecture, the loss functions are defined as follows.
First, for the entropy term, the temperature parameter $\alpha$ is tuned while learning to minimize the loss as
\begin{equation}
	\label{Eq: entropy_loss}
	\mathcal{L}(\alpha) = \sum\limits_{a_t \in \mathcal{A}} \pi (a_t|s_t) [-a_t \log \pi (a_t | s_t)] - \overline{\mathcal{H}},
\end{equation}
where $\overline{\mathcal{H}}$ denotes the target entropy that controls the desired level of exploration. Second, for the dual Q-network structure in the critic, the networks are trained to estimate the Q-value for a given state-action pair. The loss function for each Q-network $\phi_i$ is defined based on the bellman residual
\begin{equation}
    \label{Eq: critic_update}
	\mathcal{L}(\phi_i, \mathcal{D}) = \mathbb{E}_{\{s_t, a_t, s_{t+1}, r_t\} \sim \mathcal{D}} [(Q_{\phi_i}(s_t, a_t) - y_t)^2],
\end{equation}
where ${\{s_t, a_t, s_{t+1}, r_t\}}$ is sampled from the replay buffer $\mathcal{D}$, and $y_t$ is the corresponding target value computed using the target network. Third, the actor network approximates the agent's policy to determine the probability of an action for a given state. It is trained to maximize the expected Q-value while incorporating entropy regularization, formulated as follows
\begin{align}
    \label{Eq: actor_update}
	& \mathcal{L}(\psi, \mathcal{D}) = \nonumber \\
	&  \mathbb{E}_{s_t \sim \mathcal{D}} \Big[\sum_{a_t \in \mathcal{A}} \pi_{\psi}(a_t|s_t)  \Big( \alpha \log \pi_{\psi}(a_t | s_t) \! - \! \min_{i\in\{1,2\}} Q_{\phi_i}(s_t, a_t)\Big) \Big],
\end{align}
where the exploration (via the entropy term) and exploitation (via the Q-value) are balanced for action determination.

Finally, the target Q-networks with $i = 1, 2$ will be updated via soft update, that is,
\begin{equation}
	\label{Eq: soft_update}
	\hat{\phi}_i = \tau \phi_i + (1 - \tau) \hat{\phi}_i,
\end{equation}
where $\tau$ is a factor that determines the update rate for the target network parameters.

{\footnotesize
	\begin{algorithm}
		\caption{LLM Policy Distillation in LLM-HeMARL Approach.}
		\label{Alg:LLM-HeMARL_knowledge_dis}
		\DontPrintSemicolon
		\SetAlgoLined
		\KwIn{\textbf{LLM policy dataset $\mathcal{D}_{\text{LLM}}$}, initial policy $\psi$, Q-networks $\phi_{i}$ and target Q-networks $\hat{\phi}_{i}$.}
		\textbf{Initialization}: Set the maximum numbers of network updates $N_{\text{upd}}$, the iteration index $\iota=1$, $\psi_{\text{dis}}$, $\phi_{i,\text{dis}}$ and $\hat{\phi}_{i, \text{dis}}$;
		
		\For{$\iota$ \KwTo  $N_{\text{upd}}$}{
			\For{each UAV $k \in \mathcal{K}$}{
				Sample a mini-batch from $\mathcal{D}^k_{\text{LLM}}$;
				
				Update the critic networks $\phi^k_{i,\text{dis}}$ and actor network $\psi^k_{\text{dis}}$ by~\eqref{Eq: offline_dis_Q_update} and~\eqref{Eq: offline_dis_policy_update}, respectively;
				
				Soft update target critic networks $\hat{\phi}_{i, \text{dis}}$ based on~\eqref{Eq: soft_update}.
			}	
		}
		\Return{} Heuristic UAV distillation policy $\psi_{\text{dis}}$, Q-networks $\phi_{i,\text{dis}}$ and target Q-networks $\hat{\phi}_{i, \text{dis}}$, $i \in \{1, 2\}$.
	\end{algorithm}
}

{\footnotesize
	\begin{algorithm}
		\caption{Online Policy Adaptation in LLM-HeMARL Approach.}
		\label{Alg:LLM-HeMARL_on_policy_ada}
		\DontPrintSemicolon
		\SetAlgoLined
		\KwIn{\textbf{Heuristic UAV distillation policy $\psi_{\text{dis}}$, Q-networks $\phi_{i,\text{dis}}$ and target Q-networks $\hat{\phi}_{i, \text{dis}}$, $i \in \{1, 2\}$}.}
		\textbf{Initialization}: Set the maximum numbers of episodes $N_{\text{epi}}$, episode length $N_{\mathcal{T}}$ and online replay buffer $\mathcal{D}_{\text{on}}$;
		
		Load the distilled model to initialize the online model, $\psi^{(0)}_{\text{on}} = \psi_\text{dis}$, $\phi^{(0)}_{i,\text{on}} = \phi_{i, \text{dis}}$, $\hat{\phi}^{(0)}_{i, \text{on}} = \hat{\phi}_{i, \text{dis}}$, $i \in \{1, 2\}$;
		
		\For{$episode = 0$ \KwTo $N_{\text{epi}} - 1$}{
			Reset environment and set initial state $s_0$;
			
			\For{$t = 1$ \KwTo $N_{\mathcal{T}}$}{
				\For{each UAV $k \in \mathcal{K}$}{
					Sample an action $a^k_t \sim \psi^k_\text{on}(\cdot | s^k_t)$;
					
					Update UAV $k$ position $u_k(t)$;
					
					Update association status and channels;
					
					Input the channels into \textbf{Algorithm~\ref{Alg:iterative-algo}} to calculate the secrecy precoding to obtain the secrecy rate;
					
					Calculate reward $r^k(s^k_t, a^k_t)$ based on the secrecy rate and propulsion energy consumption;
					
					Store $(s^k_t, a^k_t, r^k(s^k_t, a^k_t), s^k_{t+1})$ into $\mathcal{D}^k_{\text{on}}$
					
					\If{$|\mathcal{D}| \geq |\mathcal{B}|$}{
						Sample a mini-batch $\mathcal{B}$ from $\mathcal{D}^k_{\text{on}}$;
						
						Update the critic networks, actor network, and adjust entropy temperature based on~\eqref{Eq: online_ada_Q_update},~\eqref{Eq: online_ada_policy_update} and~\eqref{Eq: entropy_loss}, respectively;
					}
					
					Soft update target critic networks $\hat{\phi}^k_{i, \text{on}}, i \in \{1,2\}$  based on~\eqref{Eq: soft_update}.
				}
			}
		}
		
		\Return{} Optimized UAV policy $\psi_{\text{on}}$.
		
	\end{algorithm}
}

\subsubsection{Policy Distillation via Offline RL}
However, offline RL often suffers from action distribution shift~\cite{kumar2019stabilizing}, causing inaccurate Q-value estimation on out-of-distribution (OOD) state-action pairs. We therefore adopt conservative Q-learning (CQL)~\cite{kumar2020conservative} to penalize OOD actions, yielding the Q-network loss
\begin{align}
    \label{Eq: CQL_update}
	& \mathcal{L}_{\text{dis}}(\phi_i, \mathcal{D}_{\text{LLM}}) = \mathcal{L}(\phi_i, \mathcal{D}_{\text{LLM}}) \nonumber \\
	& + \beta \mathbb{E}_{s_t \sim \mathcal{D}_{\text{LLM}}}\Big[\log\sum_{a_t}\exp(Q(s_t,a_t)) \! - \! \mathbb{E}_{a_t \sim \pi_{\text{LLM}}}[Q(s_t, a_t)]\Big],
\end{align}
where $\beta$ is used to control the intensity of the penalty and $\pi_{\text{LLM}}$ is the behavior policy of LLM. Then, based on~\eqref{Eq: CQL_update},  we can obtain the update method of Q network of UAV $k$ as
\begin{equation}
	\label{Eq: offline_dis_Q_update}
	\phi_{i,\text{dis}}^{(\iota+1)} \leftarrow \arg\min_{\phi_{i,\text{dis}}} \mathcal{L}_{\text{dis}}(\phi_{i,\text{dis}}, \mathcal{D}_{\text{LLM}}), i \in \{1, 2\}.
\end{equation}
Based on~\eqref{Eq: actor_update}, the actor network is updated as
\begin{equation}
	\label{Eq: offline_dis_policy_update}
	\psi^{(\iota+1)}_{\text{dis}} \leftarrow \arg\min_{\psi_{\text{dis}}} \mathcal{L}_{\text{dis}}(\psi_{\text{dis}}, \mathcal{D}_{\text{LLM}}).
\end{equation}
As such, the algorithmic process of policy distillation is summarized in Algorithm~\ref{Alg:LLM-HeMARL_knowledge_dis}.

\subsubsection{Online Adaptation via Online RL}
This step further adapts the distilled agents to the deployment environment. We initialize the online model with the distilled model from Algorithm~\ref{Alg:LLM-HeMARL_knowledge_dis} and grant exploration ability by tuning the entropy temperature $\alpha$. During each episode, every UAV $k$ selects an action from $\psi^k_{\text{on}}(\cdot | s^k_t)$, updates its position and channels, and obtains the secrecy rate via Algorithm~\ref{Alg:iterative-algo}; the resulting transition $(s^k_t, a^k_t, r^k_t, s^k_{t+1})$ is stored in the replay buffer $\mathcal{D}^k_{\text{on}}$. Once $|\mathcal{D}| \geq |\mathcal{B}|$, the networks are updated by minimizing their losses over sampled mini-batches. Based on~\eqref{Eq: critic_update}, the critic networks are updated as
\begin{equation}
	\label{Eq: online_ada_Q_update}
	\phi_{i, \text{on}}^{\iota+1} \leftarrow \arg\min_{\phi_i} \mathcal{L}_{\text{on}}(\phi_{i,\text{on}}, \mathcal{D}_{\text{on}}), i \in \{1,2\}.
\end{equation}
The actor network update method based on~\eqref{Eq: actor_update} is
\begin{equation}
	\label{Eq: online_ada_policy_update}
	\psi^{(\iota+1)}_{\text{on}} \leftarrow \arg \min_{\psi_{\text{on}}} \mathcal{L}_{\text{on}} (\psi_{\text{on}}, \mathcal{D}_{\text{on}}).
\end{equation}
The parameters of the target critic networks are updated periodically using soft update rules while the predicted critics are being trained. The overall algorithm is summarized in Algorithm~\ref{Alg:LLM-HeMARL_on_policy_ada}.

\subsection{Complexity Analysis}
We analyze the complexity of the three steps of LLM-HeMARL as follows.
\begin{itemize}
\item LLM Expert Policy Collection: This step includes the reasoning of LLM and the solution of the secrecy precoding through the S2DC. Hence, the computational complexity of this step can be derived as $\mathcal{O}(N_{\text{d}}N_{\mathcal{T}}(C_{\text{LLM}} + \mathcal{O}_{\text{S2DC}}))$, where $N_d$ is the number of episode policies to be collected.

\item LLM Policy Distillation: According to Algorithm~\ref{Alg:LLM-HeMARL_knowledge_dis}, the computational complexity of this step is estimated to be $\mathcal{O}(N_{\text{upd}} N_{\mathcal{K}}(2|\phi| + |\psi|))$, where $|\phi|$ and $|\psi|$ are the numbers of parameters of the critic and actor networks, respectively.
\item Online Policy Adaptation: The computational complexity of this step mainly comes from the environment interaction, S2DC and network updates, so it can be summarized as $\mathcal{O}(N_{\text{epi}}N_{\mathcal{T}}N_{\mathcal{K}}(|\psi| + \mathcal{O}_{\text{S2DC}} + |\mathcal{B}|(|\psi| + 2|\phi|)))$.
\end{itemize}

It is worth noting that the main computational overhead of the proposed framework stems from the high latency inference of LLM $\mathcal{O}(C_{\text{LLM}})$. However, as the LLM-generated heuristic expert policies are precomputed and used to guide the learning process rather than being directly involved in real-time decision-making for precoding and trajectory optimization. As a result, the proposed approach is able to almost meet the stringent latency requirements of practical communication systems.

\section{Performance Evaluation}
In this section, we evaluate the performance of the proposed approach in secure HetUAVNs. All experiments were carried out on a computer host equipped with an AMD EPYC 9654 CPU and a NVIDIA GeForce RTX 4080 GPU. We used PyTorch 2.2.2 for deep learning implementations and the MOSEK solver for convex optimization tasks.

\subsection{Simulation Settings}
\emph{Parameter Settings:}
For the LLM component, we use DeepSeek-R1~\cite{guo2025deepseek} via its public APIs, with temperature $0.0$  for deterministic outputs, $top_p=0.95$, and a $max_tokens$ limit of $16384$ to accommodate its chain-of-thought reasoning. For the MOSEK solver, the relative optimality gap tolerance is set to $10^{-3}$, with $32$ CPU threads  for parallel computation.

To better capture the large-scale input features of GTs and Eves, the proposed LLM-HeMARL adopts the actor and critic networks consisting of a three-layer Transformer encoder, followed by three fully connected layers (with 256, 256, 128 neurons, respectively) and ReLU activation functions, as shown in Fig.~\ref{Fig:algo_framework}. The Transformer encoder uses a model dimension of $64$ and $4$ attention heads. The learning rate of each actor network and critic network is set to $5 \times 10^{-4}$ and the discount factor is set to $\gamma=0.99$. The distillation and online adaptation processes are run for $N_{\text{upd}}=500$ and $N_{\text{epi}}=5000$ episodes, respectively, with corresponding mini-batch sizes of $\mathcal{B}=512$ and $1024$.
The LLM policy dataset $\mathcal{D}_{\text{LLM}}$ was constructed by collecting $10,000$ state-action-reward transition samples through $N_d=250$ rollout episodes. Environment-related parameters are summarized in Table~\ref{Tab:table-env-para}.

\emph{Baseline Settings:} To comprehensively evaluate the performance of the proposed method in secure HetUAVNs, we compare it against five baseline approaches, described as follows:
\begin{itemize}
\item LLM-HeMARL-S2DC (Ours): The proposed approach in this work.
\item ISAC-S2DC: This combines ISAC for UAV trajectories optimization without LLM expert policy guidance and S2DC for secrecy precoding.
\item ISAC: This baseline applies the ISAC to jointly optimize both UAV trajectories and secrecy precoding.
\item MASAC-S2DC: A multi-agent SAC variant from \cite{li2023computation} to solve trajectories, with shared replay buffer across agents, combined with S2DC for secrecy precoding.
\item MASAC: This baseline uses the MASAC to solve both trajectories and secrecy precoding jointly.
\item SCA-S2DC: This baseline uses the successive convex approximation (SCA) method to optimize UAV trajectories, and applies S2DC for secrecy precoding.
\end{itemize}

Each baseline is evaluated using multiple random seeds $[30,40,50,60]$ to assess robustness and generalization. To be fair, all approaches are run with the aforementioned parameters and use the same actor and critic networks structure.

\begin{table}[t]
	\renewcommand{\arraystretch}{1.5}
	\caption{\textsc{Parameters Settings.}}
	\label{Tab:table-env-para}
	\centering
	\resizebox{0.9\linewidth}{!}{
		\begin{tabular}{cc}
			\Xhline{1px}
			\textbf{Parameters} & \textbf{Values (Unit)}\\
			\Xhline{1px}
			Maximum and minimum velocity of UAV ($v_{\text{max}}$,$v_{\text{min}}$) & 25, 4 (m/s) \\
			\hline
			Flight altitude of UAV ($H_{\text{UAV}}$) & 100 m \\
			\hline
			Central carrier frequency ($f_c$) & 2.4 (GHz) \\
			\hline
			Maximum power of UAV ($P_{\text{max}}$) & 35 (w) \\
			\hline
			PSD of AWGN at GTs ($\sigma^2$) & -170 (dBm/Hz) \\
			\hline
			Channel S-curve parameters ($\delta$, $f$) & 9.61, 0.15 \\
			\hline
			Excessive path loss exponent ($\eta_{\text{LoS}}$, $\eta_{\text{NLoS}}$) & 1, 20 (dB) \\
			\hline
			The number of antennas ($M$) & 2 \\
			\hline
			Fuselage drag ratio ($d_0$) & 0.3 \\
			\hline
			Air density ($\rho_a$) & 1.225 \\
			\hline
			Rotor solidity ($s_\text{sol}$) & 0.05 \\
			\hline
			Rotor disc area ($A$) & 0.503 \\
			\hline
			Speed of the rotor blade ($v_{\text{tip}}$) & 120 \\
			\hline
			Safe distance between UAVs ($d_c$) & 5 m \\
			\Xhline{1px}
	\end{tabular}}
\end{table}

\subsection{Discussion on the Effectiveness of LLM Expert Policy}
\label{sec:discussion}

To evaluate the effectiveness of the expert policy generated by LLMs in handling the complex cooperation problem of heterogeneous UAV networks, we designed a controlled experiment that directly deploys an LLM as a trajectory planner within the proposed system environment.

Consider a $400\,\text{m}\times400\,\text{m}$ area with $N_{\mathcal{I}}=32$ GTs randomly distributed around hot spots and $N_{\mathcal{E}}=5$ Eves randomly distributed throughout the area. Four UAVs are initialized at positions $\{[175,175],[225,225],[175,225],[225,175]\}$, centered within the area. To reflect UAV heterogeneity, their coverage ranges are set to $[50,75,25,50]$ meters, respectively, with corresponding service capacities of $[5,7,3,5]$ GTs simultaneously.

\begin{figure}[!t]
	\centering
	\includegraphics[width=0.3\textwidth]{"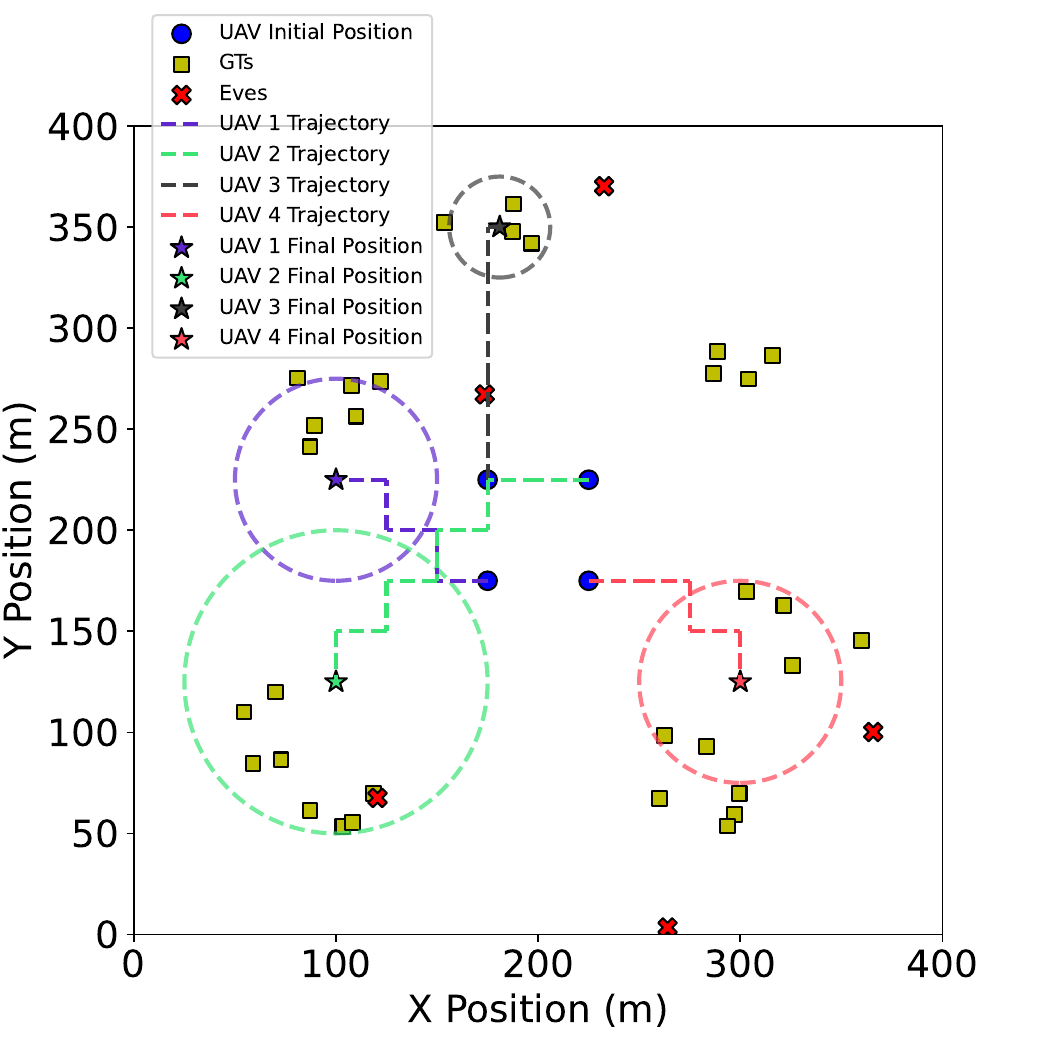"}
	\caption{UAV trajectories generated by the LLM expert policy in the heterogeneous UAV network.}
	\label{Fig:traj_llm}
\end{figure}

As illustrated in Fig.~\ref{Fig:traj_llm}, the LLM expert policy demonstrates a clear global awareness of UAV heterogeneity, intelligently allocating tasks according to each UAV's distinct capabilities and thereby achieving an effective division of labor and cooperation. Taking UAV~2 as a representative example: with the largest service capacity of $7$ and coverage range of $75$ m, it is the most capable UAV in the swarm. Under a rule-based greedy strategy, UAV~2 would only associate with the $5$ nearest GTs, severely underutilizing its coverage advantage. In contrast, under LLM guidance, UAV~2 actively patrols a broader region that better matches its coverage range, exploiting its capacity advantage to serve a wider and denser cluster of GTs. This observation confirms that LLMs can reason about heterogeneous agent capabilities and translate such reasoning into effective coordination strategies, which motivates their use as an expert policy source in the proposed LLM-HeMARL framework.

\subsection{Performance Results}
\subsubsection{Convergence Analyses and Comparisons}

\begin{figure}[!t]
	\centering
	\includegraphics[width=0.3\textwidth]{"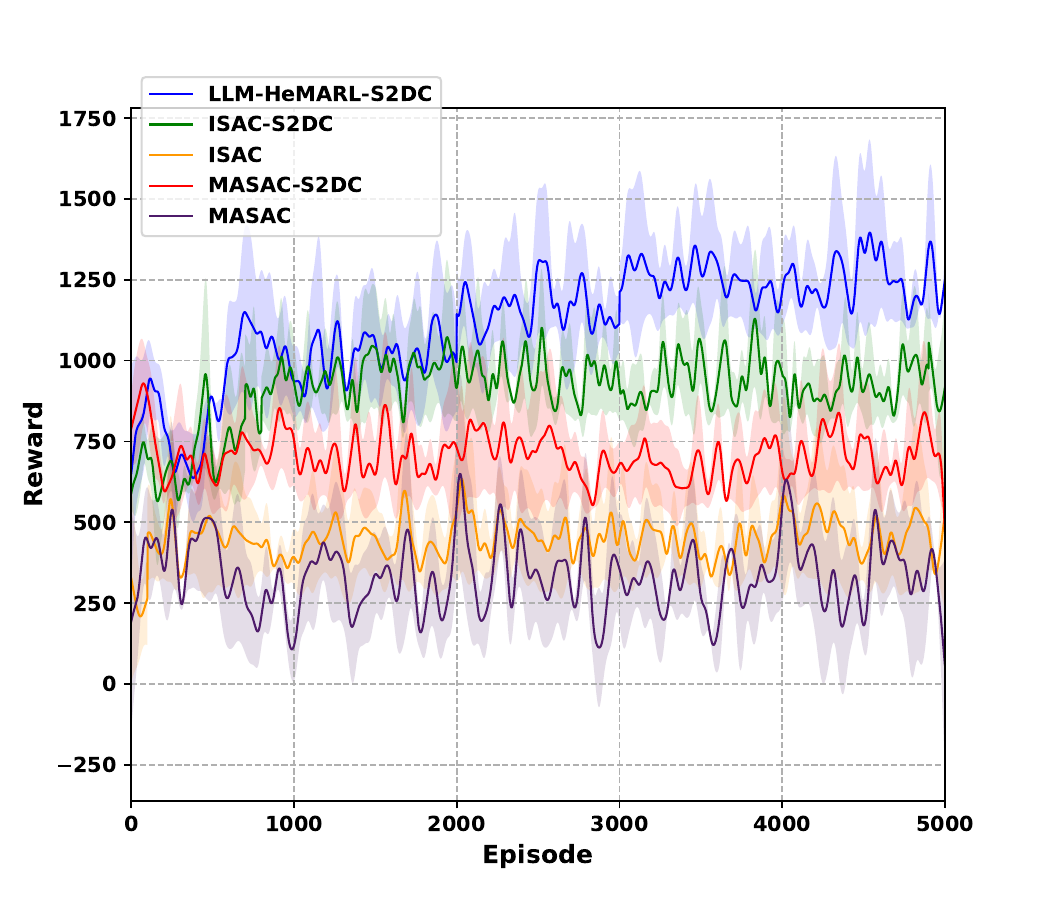"}
	\caption{Convergence comparison of the baselines over training.}
	\label{Fig:train_returns}
\end{figure}

To compare the convergence performance of the algorithms, we will continue to use the network scenario described above. Each episode spans $N_{\mathcal{T}}=40$ time slots. Fig.~\ref{Fig:train_returns} compares the convergence behavior of the proposed approach with baselines in terms of episode reward under different random seeds, where the shaded area represents the variance and the solid line denotes the mean. It can be observed that, thanks to the guidance of the LLM expert policy, our method achieves a higher initial reward. After a brief decline, the agent quickly adapts to the environment, and the reward steadily increases. Compared to ISAC-S2DC, the integration of the LLM expert policy improves performance by approximately $25\%$. By comparing ISAC-S2DC and MASAC-S2DC, we verify that experience sharing in HetUAVNs may lead to performance degradation. A similar phenomenon is also seen in the comparison between ISAC and MASAC. In addition, by comparing ISAC-S2DC with ISAC or MASAC-S2DC with MASAC, we find that the hierarchical optimization framework effectively decouples complex problems and greatly improves the performance of the algorithm. It is also worth noting that all baselines exhibit oscillations due to dynamic environmental changes and time-varying CSI. In contrast, the proposed method demonstrates superior stability and faster convergence in adapting to new environments, benefiting from its expert-guided policy initialization.

\begin{figure}[t]
	\centering
	\captionsetup{font=small}
	\begin{subfigure}[t]{0.24\textwidth}
		\includegraphics[width=\linewidth]{"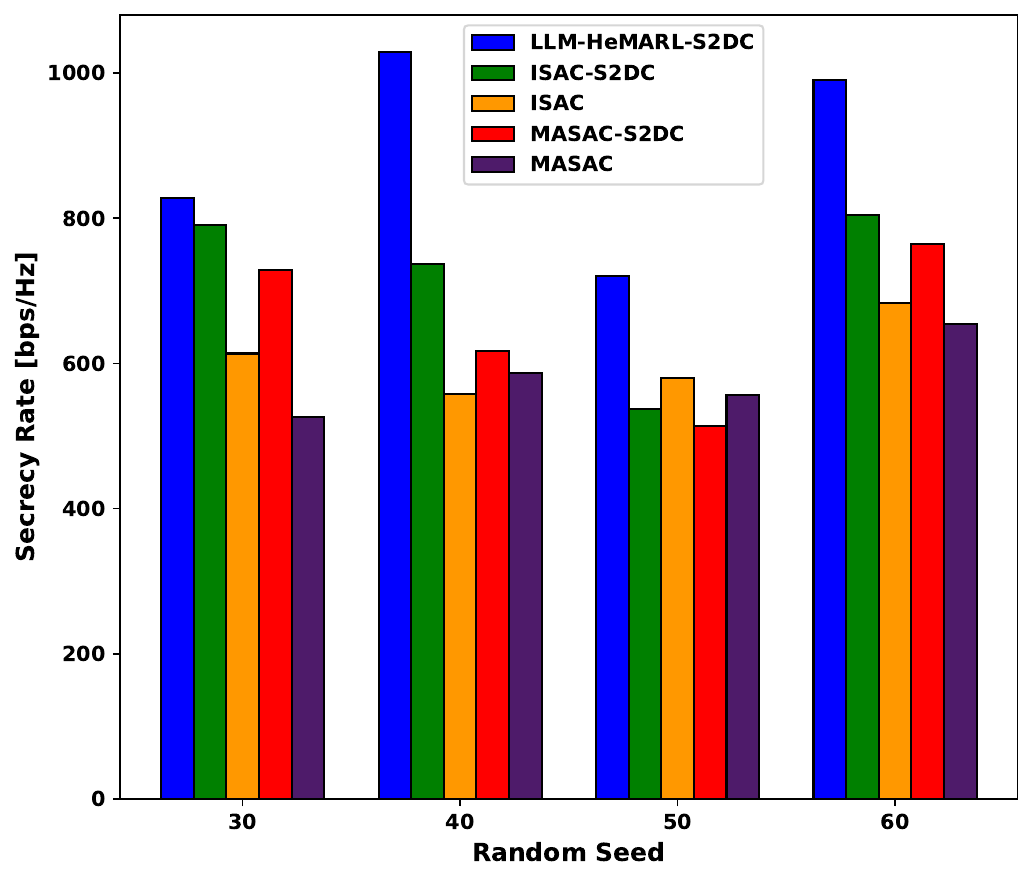"}
		\caption*{(a) Secrecy Rate}
	\end{subfigure}
	\begin{subfigure}[t]{0.24\textwidth}
		\includegraphics[width=\linewidth]{"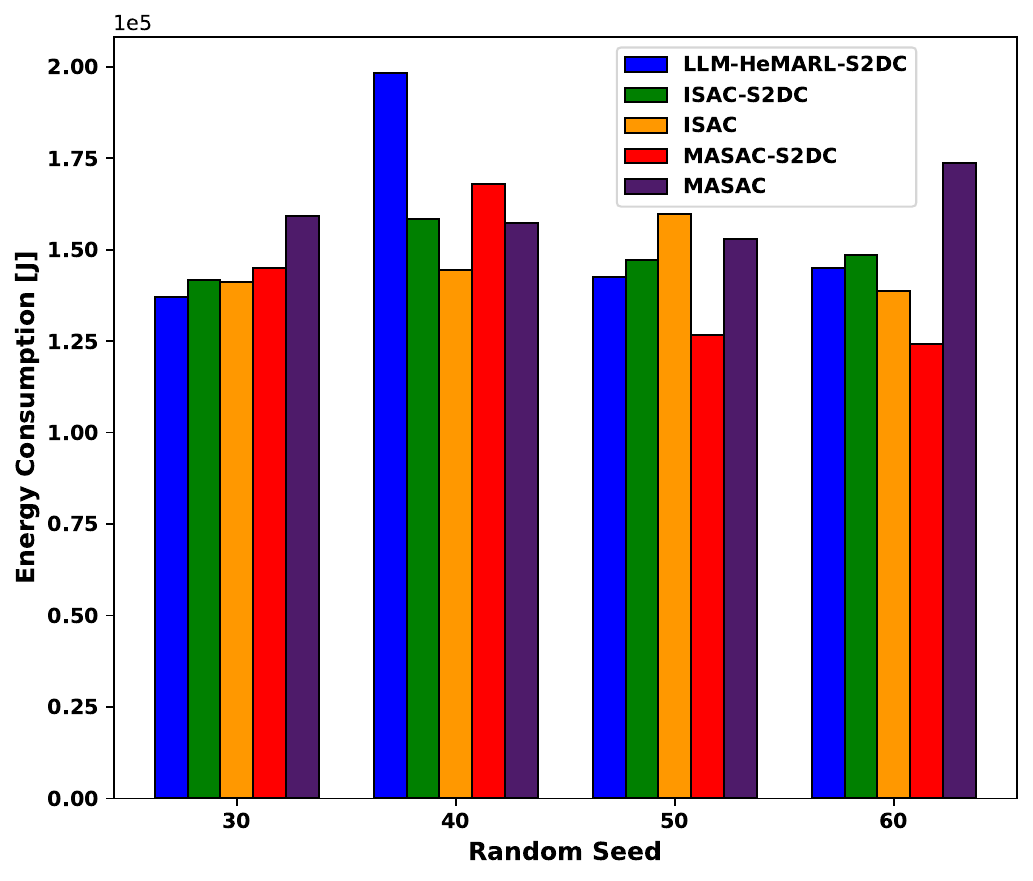"}
		\caption*{(b) Propulsion Energy Consumption}
	\end{subfigure}
	\caption{Performance comparison of the baselines under different random seeds in one episode.}
	\label{Fig:indicators_comparison}
\end{figure}

To provide a more intuitive demonstration of the proposed solution's performance when deployed, Fig.~\ref{Fig:indicators_comparison} presents a detailed comparison of five methods under different random seeds from objective $1$ (Secrecy Rate) and objective $2$ (Propulsion Energy Consumption). As shown in Fig.~\ref{Fig:indicators_comparison}(a), the hierarchical optimization framework achieves a higher secrecy rate than the coupled solution approach. Moreover, we can also find the same phenomenon that due to the fact that heterogeneity reduces the experience efficiency, the approach using the ISAC performs better than the method using the MASAC in terms of both objectives. It is worth noting that the trade-off between objectives may lead to partial preference in optimization. For instance, when the random seed is $40$, UAVs tend to sacrifice propulsion efficiency in favor of maximizing secrecy performance. Overall, compared to methods relying on coupled optimization and shared experience, the proposed approach demonstrates superior capability in identifying a better Pareto frontier within the large solution space induced by multi-objective trade-offs in HetUAVNs.

\begin{figure}[t]
	\centering
	\captionsetup{font=small}
	\begin{subfigure}[t]{0.24\textwidth}
		\includegraphics[width=\linewidth]{"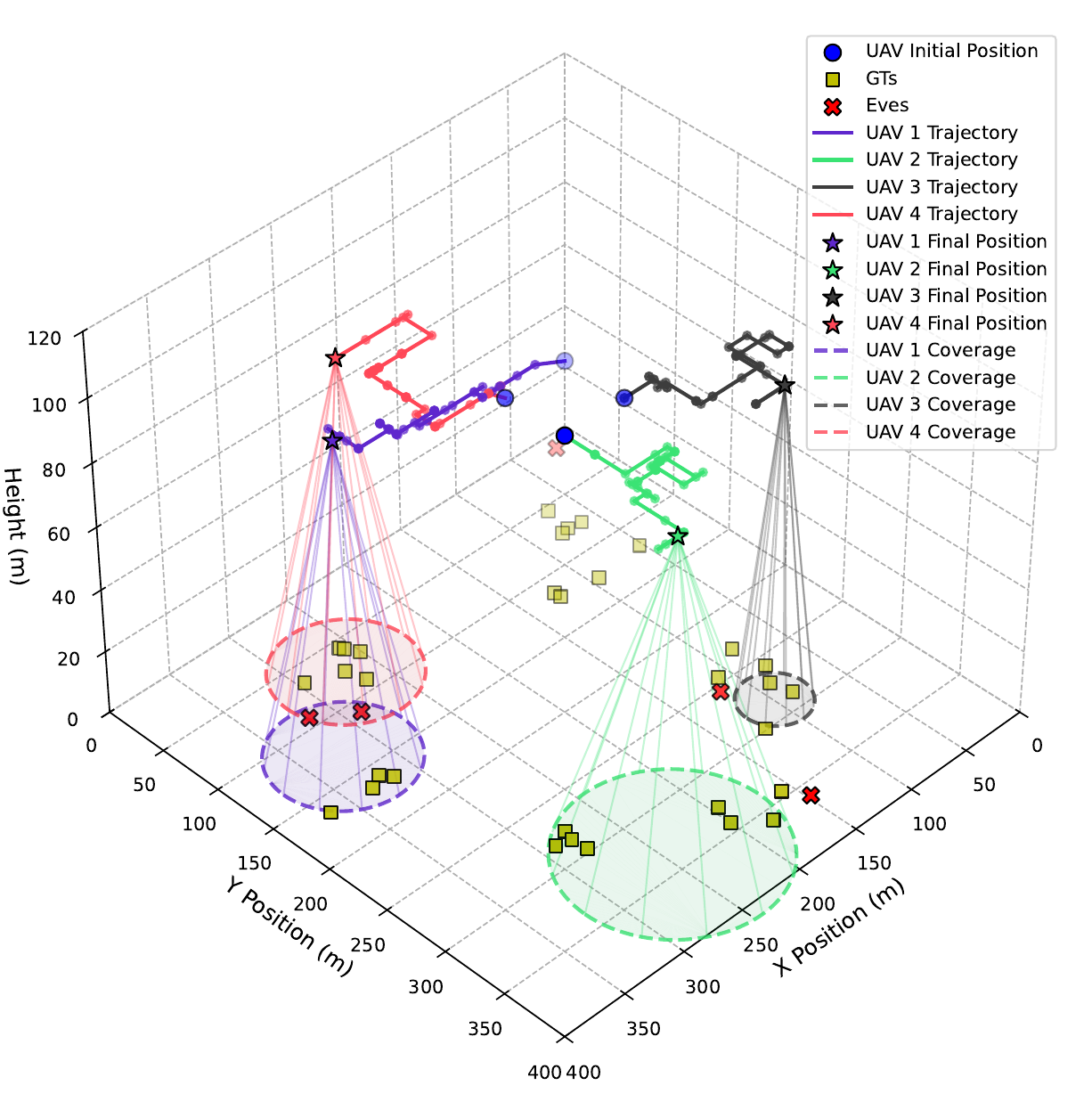"}
		\caption*{(a) $seed=30$}
	\end{subfigure}\hspace{-5pt}
	\begin{subfigure}[t]{0.24\textwidth}
		\includegraphics[width=\linewidth]{"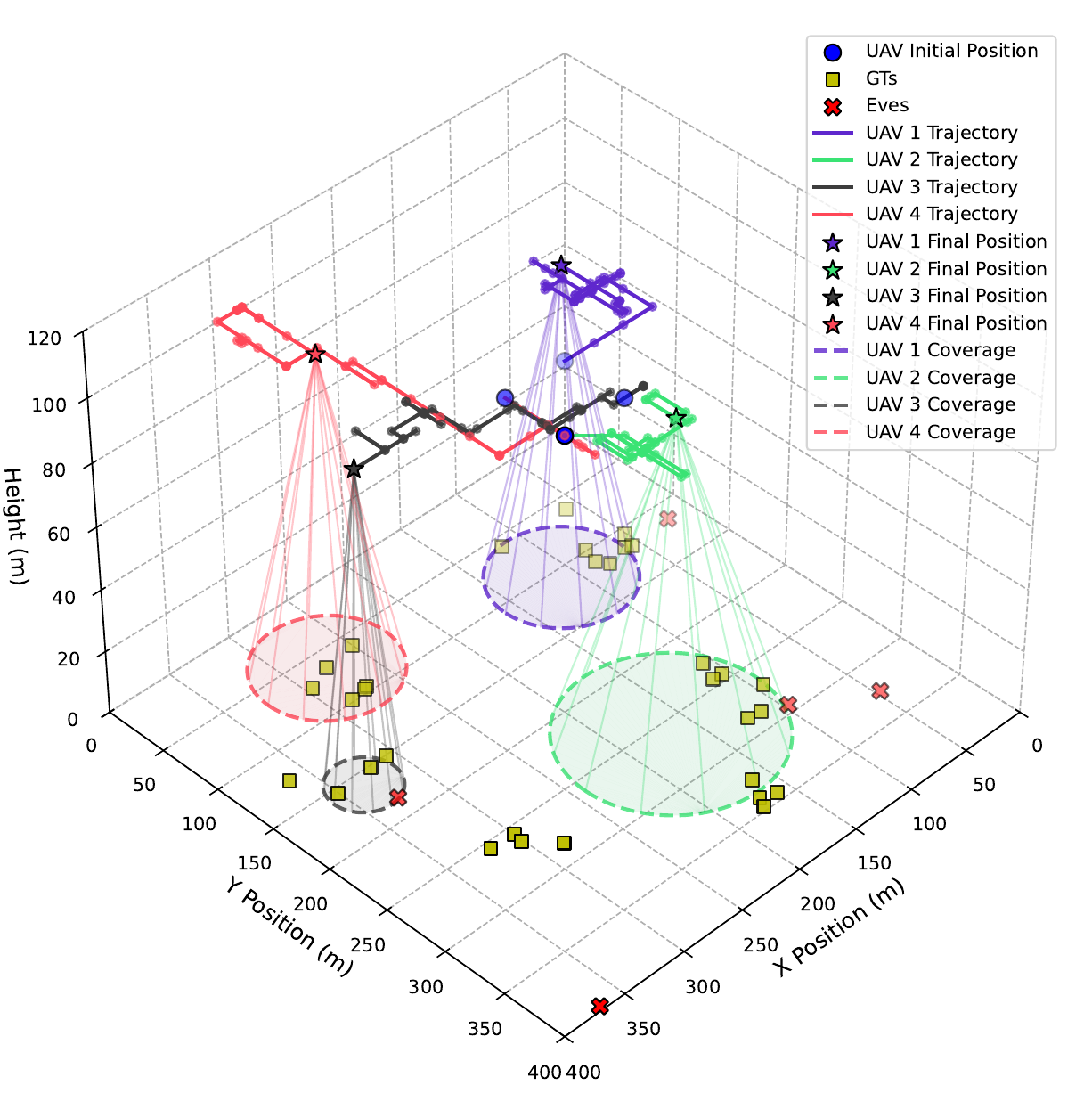"}
		\caption*{(b) $seed=40$}
	\end{subfigure}\hspace{-5pt}
	\begin{subfigure}[t]{0.24\textwidth}
		\includegraphics[width=\linewidth]{"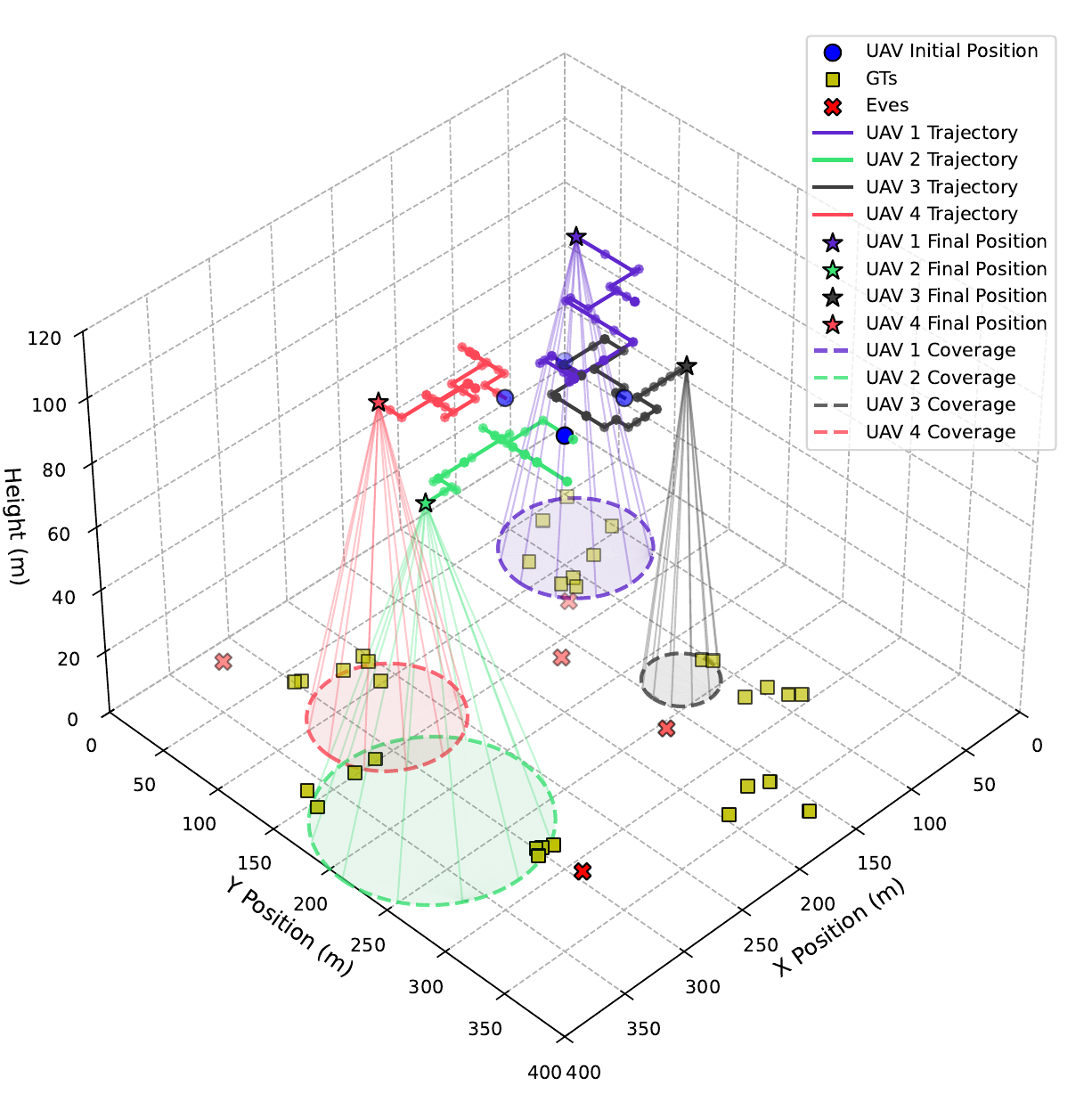"}
		\caption*{(c) $seed=50$}
	\end{subfigure}\hspace{-5pt}
	\begin{subfigure}[t]{0.24\textwidth}
		\includegraphics[width=\linewidth]{"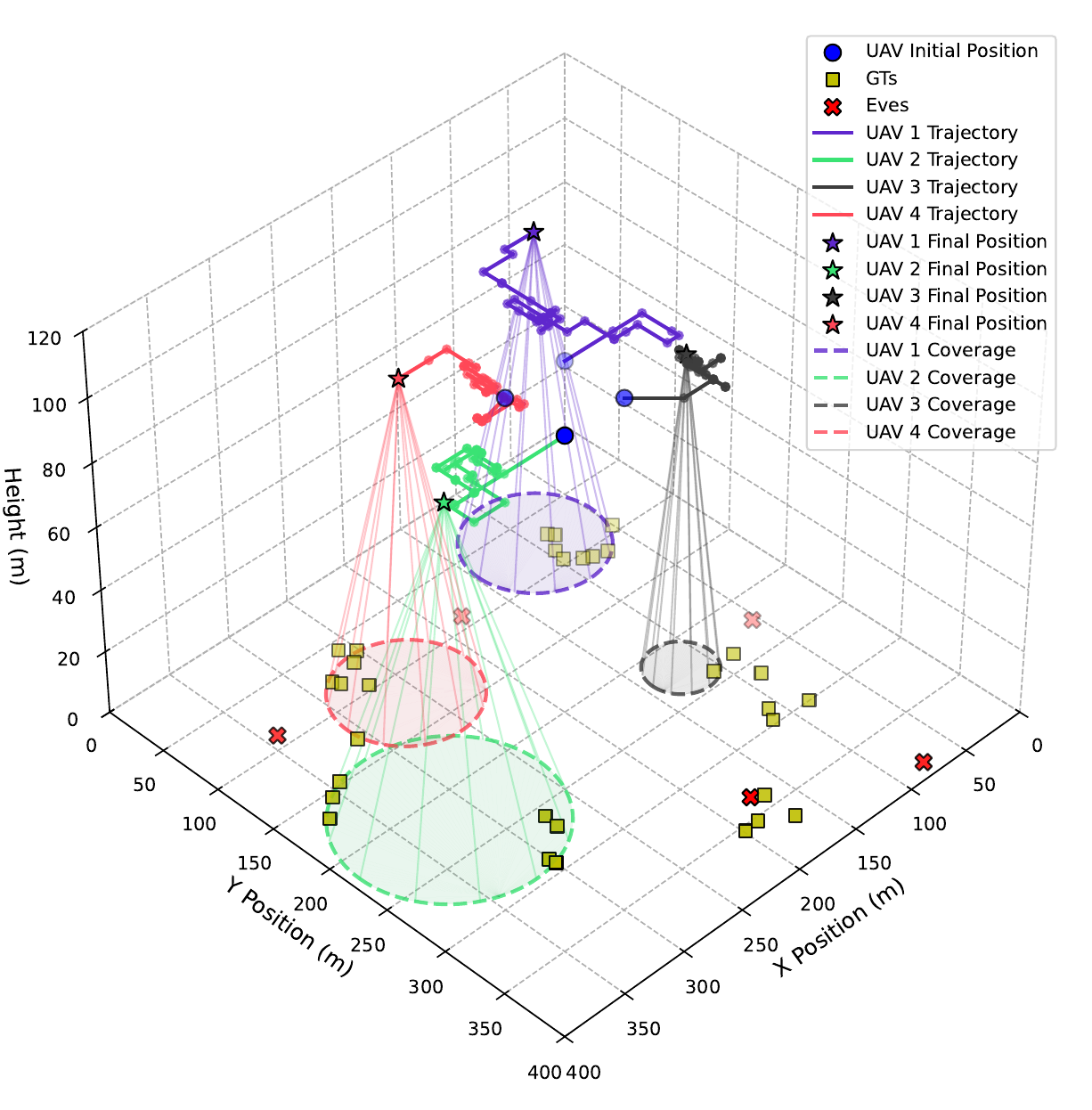"}
		\caption*{(d) $seed=60$}
	\end{subfigure}
	\caption{UAV trajectories under different random seeds in one episode.}
	\label{Fig:traj_in_different_seed}\vspace{-0.2cm}
\end{figure}

Fig.~\ref{Fig:traj_in_different_seed} illustrates the trajectories of heterogeneous UAVs over $N_{\mathcal{T}}$ time slots under different random seeds. As shown in Fig.~\ref{Fig:indicators_comparison}, when the random seed is $40$ or $60$, Eves are located farther from GT hot spots, resulting in higher secrecy rates compared to seeds $30$ and $50$. Overall, it can be observed that all UAVs effectively identify coverage positions according to their heterogeneous coverage ranges and service capabilities.

\subsubsection{Impact of Different Numbers of UAVs}
To evaluate the impact of UAV quantity on approach performance, we test the proposed method in a larger-scale scenario. Specifically, we consider an $800 \times 800$ $m^2$ square grid area, in which $100$ GTs are located. The episode length is changed to $N_{\mathcal{T}}=20$ time slots. The GT density follows a fat-tailed distribution, i.e., a majority of users cluster in a few hot spots while a minority are sparsely scattered across the rest of the area \cite{1b42e592b985443e96700757887252bf}. The number of UAVs $N_{\mathcal{K}}$ is set to $[2, 4, 6, 8, 10]$. For each UAV $k$, its coverage range $C^r_k$ is randomly sampled from $[80,120]$ meters, and its service capacity $N^s_k$ is selected from $[5, 15]$, reflecting UAV heterogeneity.

\begin{figure*}[tp]
	\centering
	\captionsetup{font=small}
	\hspace{-5pt}
	\begin{subfigure}[t]{0.25\textwidth}
		\includegraphics[width=\linewidth]{"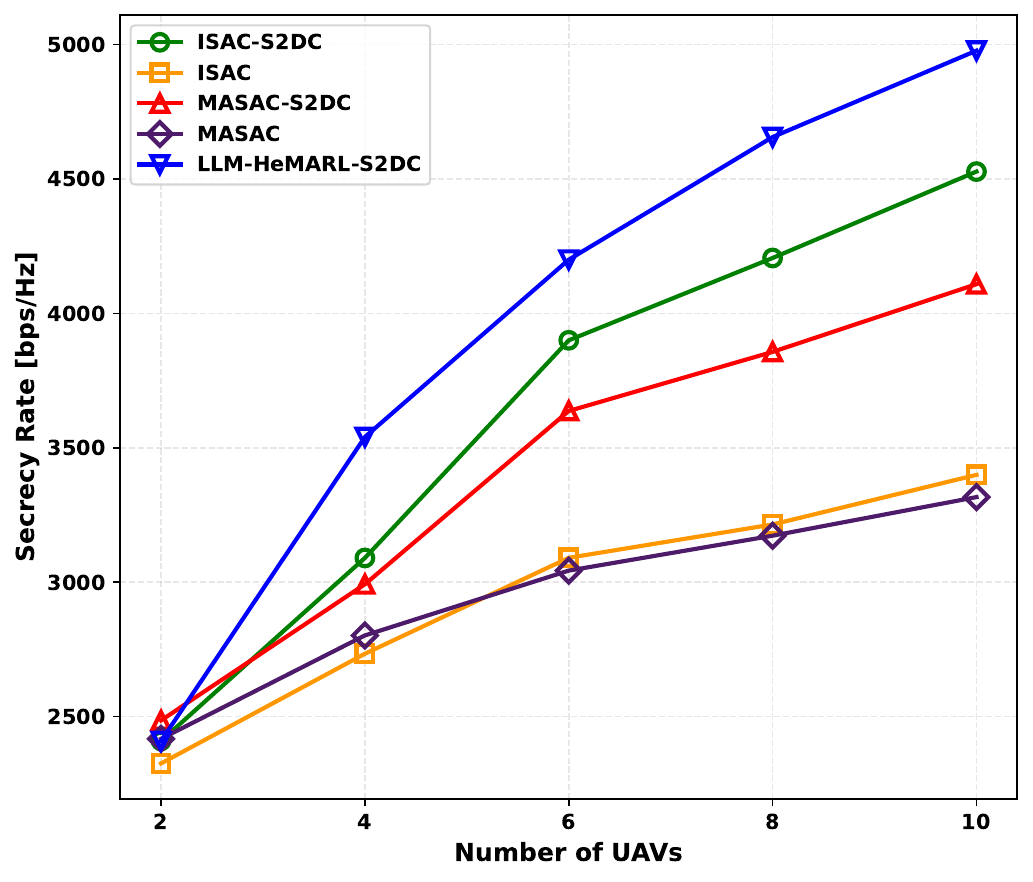"}
		\caption*{(a) $seed=30$}
	\end{subfigure}\hspace{-5pt}
	\begin{subfigure}[t]{0.25\textwidth}
		\includegraphics[width=\linewidth]{"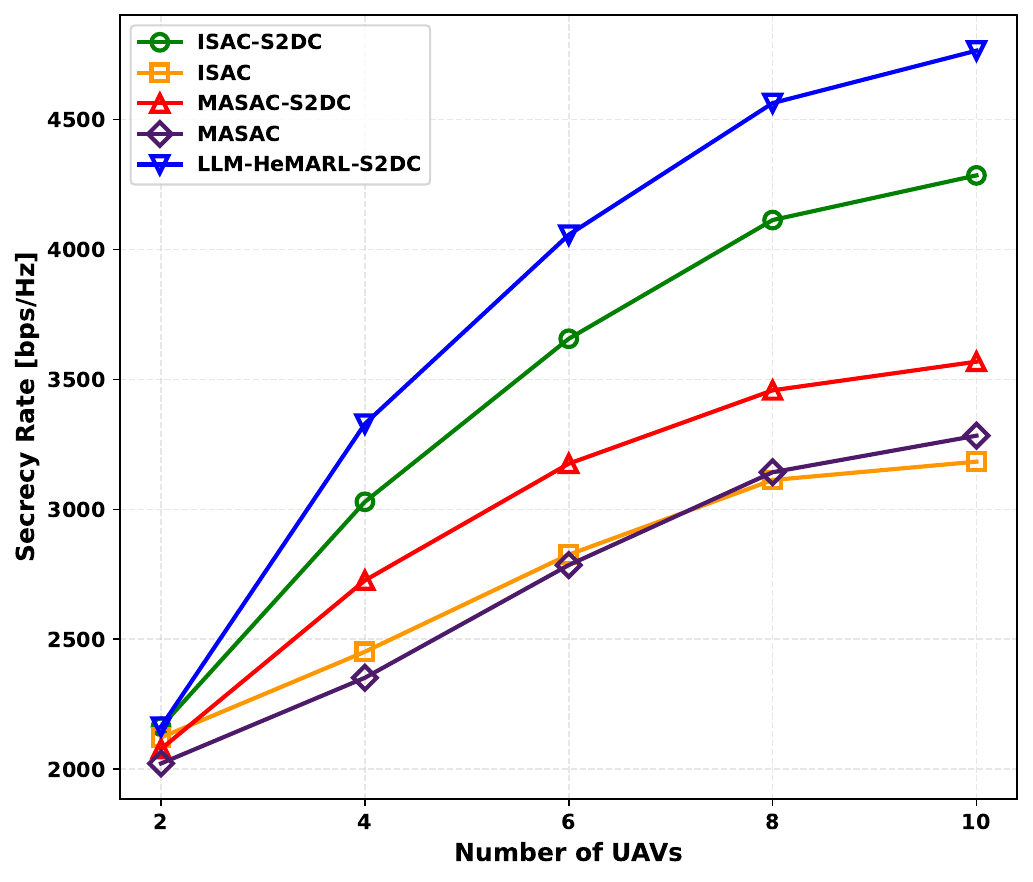"}
		\caption*{(b) $seed=40$}
	\end{subfigure}\hspace{-5pt}
	\begin{subfigure}[t]{0.25\textwidth}
		\includegraphics[width=\linewidth]{"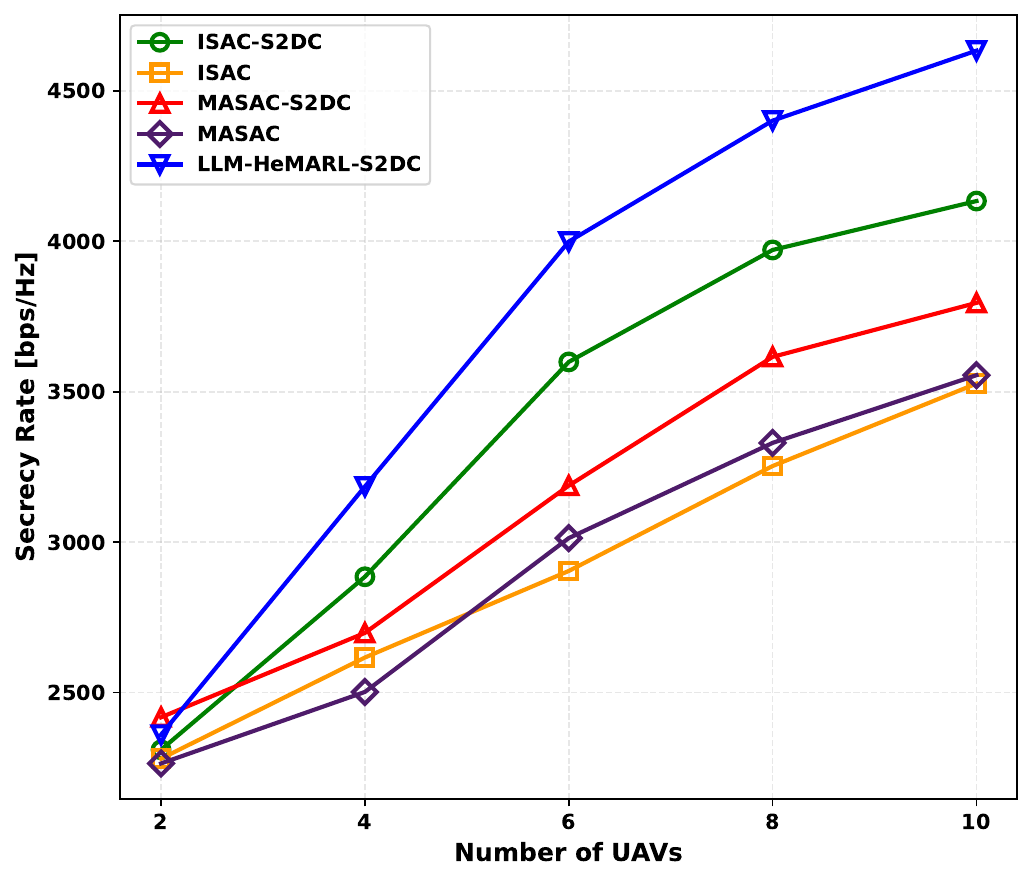"}
		\caption*{(c) $seed=50$}
	\end{subfigure}\hspace{-5pt}
	\begin{subfigure}[t]{0.25\textwidth}
		\includegraphics[width=\linewidth]{"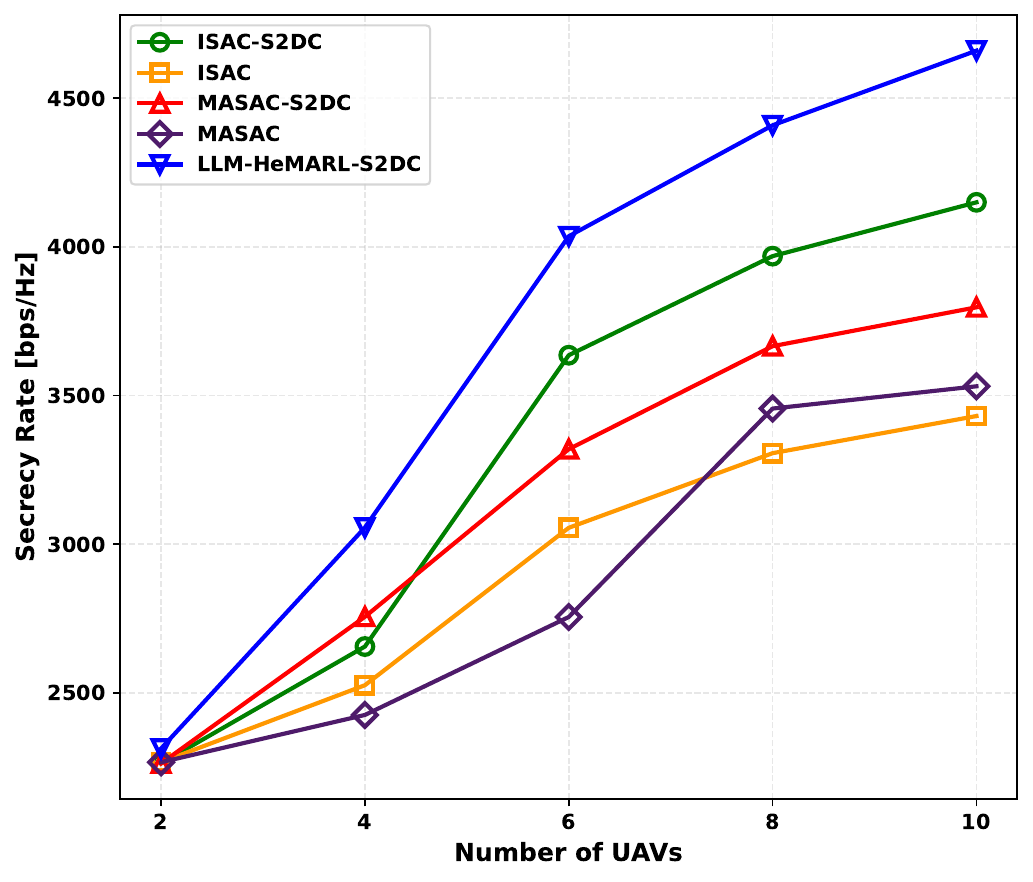"}
		\caption*{(d) $seed=60$}
	\end{subfigure}
	\caption{The cumulative secrecy rate under different numbers of UAVs in one episode.}
	\label{Fig:secrecy_in_different_seed_and_number_uav}\vspace{-0.2cm}
\end{figure*}

As can be seen from Fig.~\ref{Fig:secrecy_in_different_seed_and_number_uav}, the secrecy rate increases with the number of deployed UAVs for all baselines. Since the number of GTs is fixed, the growth rate gradually diminishes as more UAVs are added. When the number of UAVs is small, the performance gain of the proposed approach is marginal compared to baselines.
However, as the number of UAVs increases from $6$ to $8$, the secrecy performance improves by $2\% \sim 10 \%$ relative to the baseline SCA-S2DC,
demonstrating the effectiveness of the proposed method in achieving secure communication in HetUAVNs.

\begin{figure*}[t]
	\centering
	\captionsetup{font=small}
	\hspace{-5pt}
	\begin{subfigure}[t]{0.25\textwidth}
		\includegraphics[width=\linewidth]{"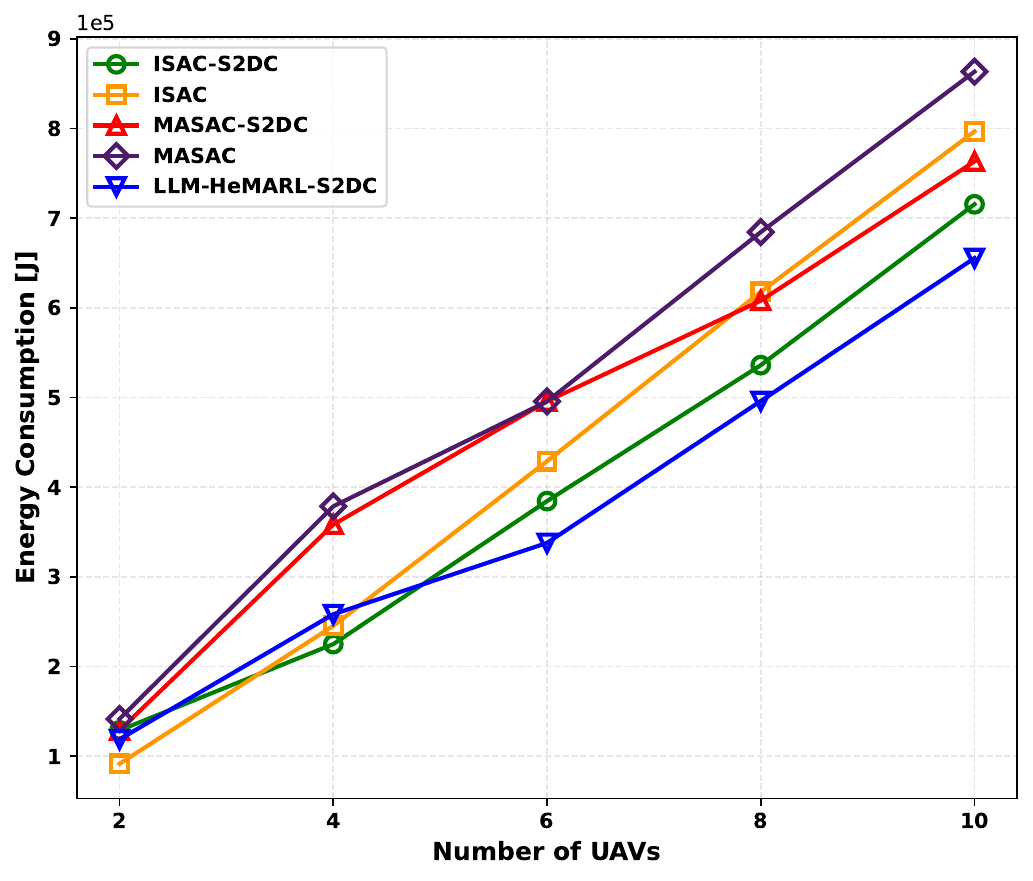"}
		\caption*{(a) $seed=30$}
	\end{subfigure}\hspace{-5pt}
	\begin{subfigure}[t]{0.25\textwidth}
		\includegraphics[width=\linewidth]{"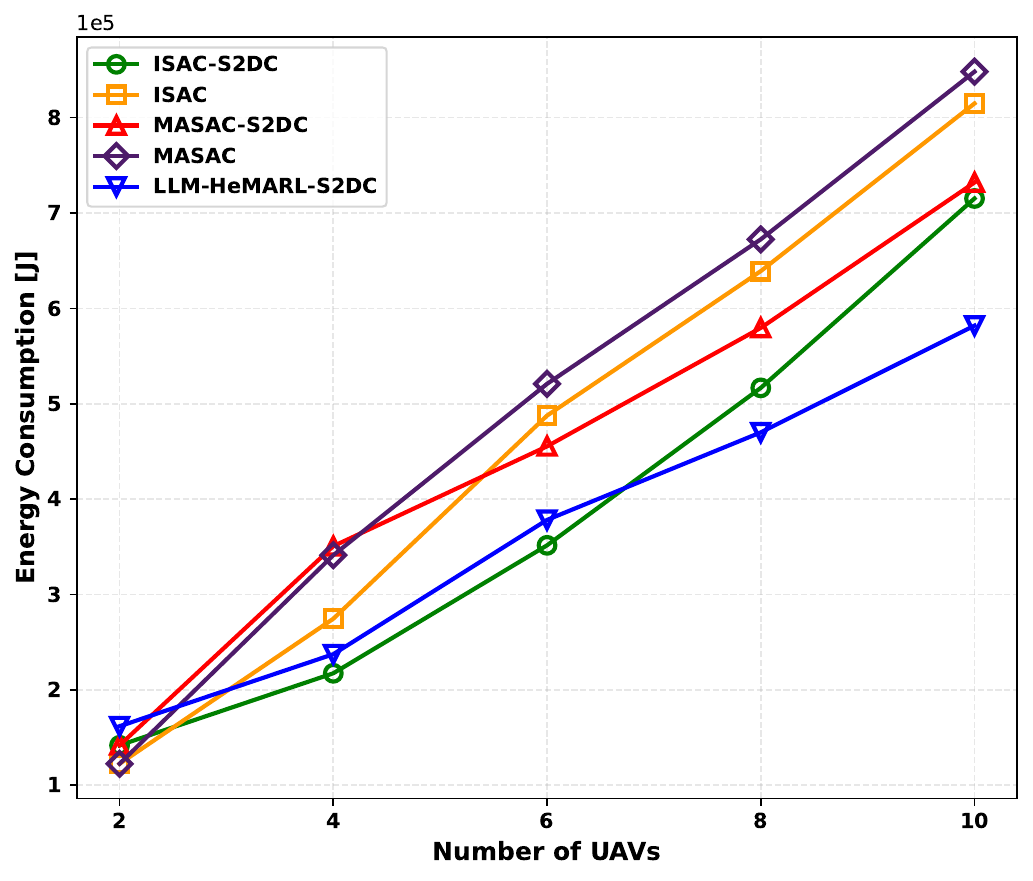"}
		\caption*{(b) $seed=40$}
	\end{subfigure}\hspace{-5pt}
	\begin{subfigure}[t]{0.25\textwidth}
		\includegraphics[width=\linewidth]{"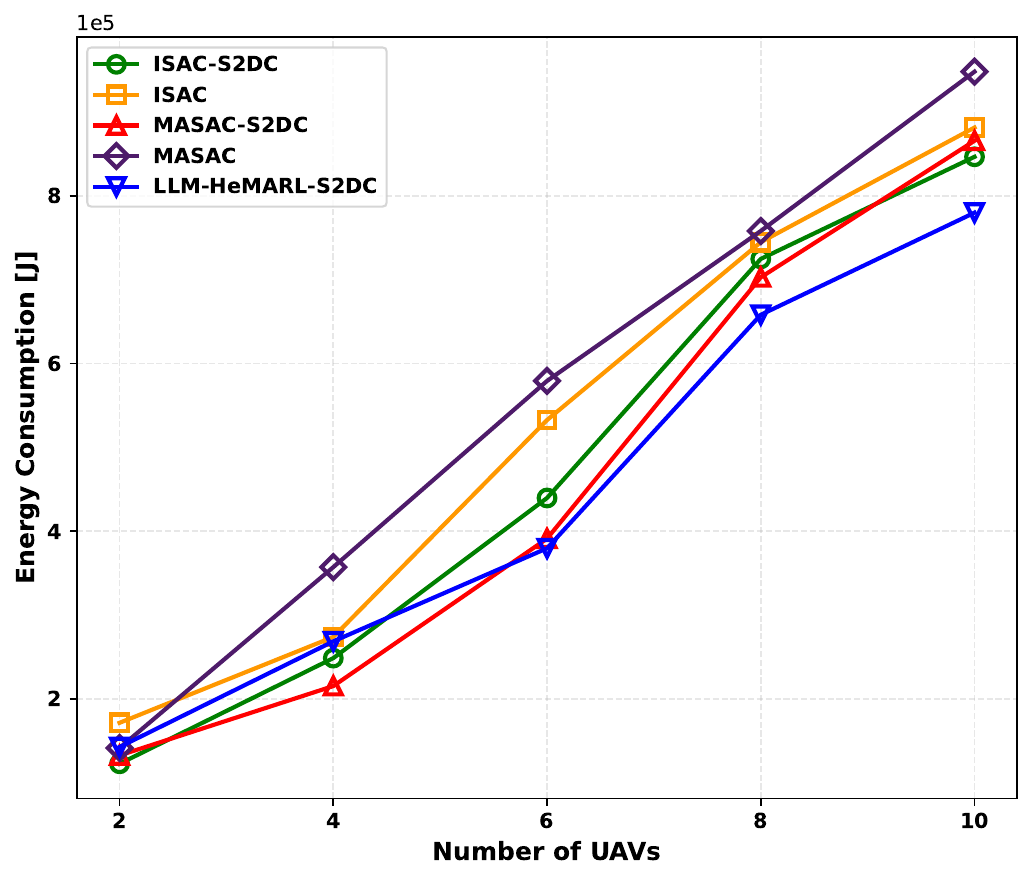"}
		\caption*{(c) $seed=50$}
	\end{subfigure}\hspace{-5pt}
	\begin{subfigure}[t]{0.25\textwidth}
		\includegraphics[width=\linewidth]{"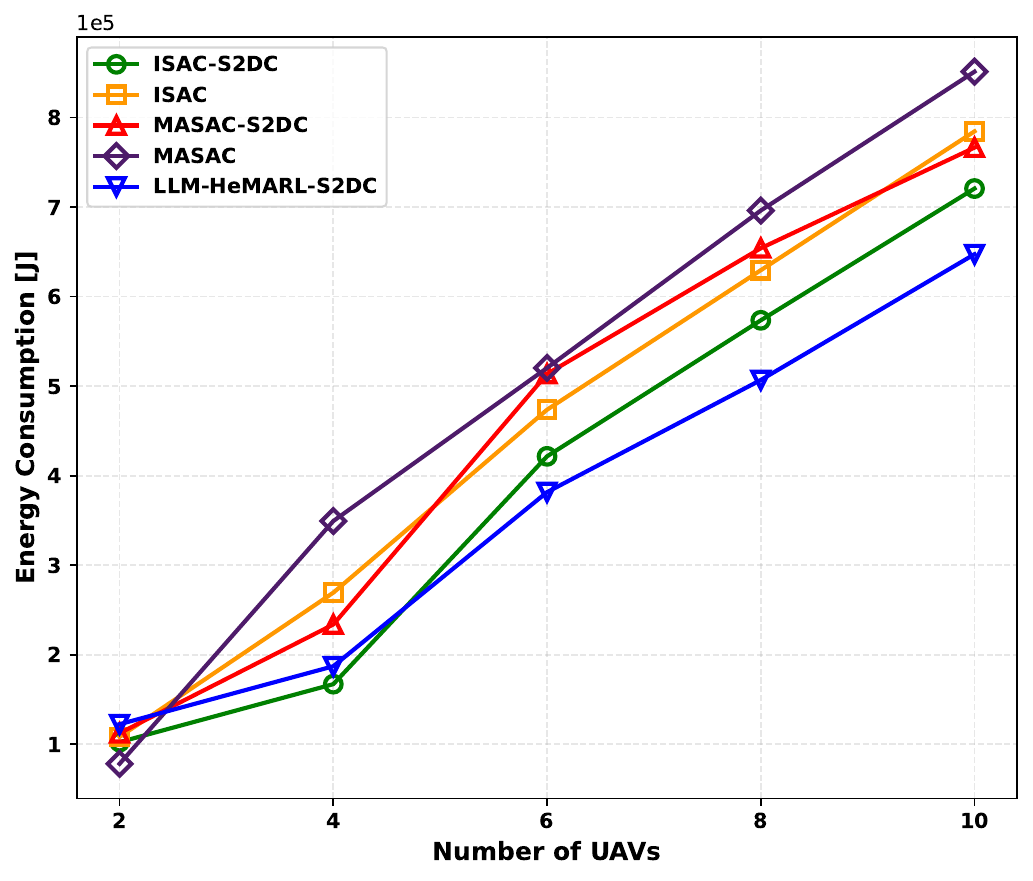"}
		\caption*{(d) $seed=60$}
	\end{subfigure}
	\caption{The cumulative propulsion energy consumption under different numbers of UAVs in one episode.}
	\label{Fig:energy_consumption_in_different_seed_and_number_uav}\vspace{-0.2cm}
\end{figure*}

Fig.~\ref{Fig:energy_consumption_in_different_seed_and_number_uav} compares the propulsion energy consumption of different approaches under varying numbers of UAVs. Unlike the secrecy rate shown in Fig.~\ref{Fig:secrecy_in_different_seed_and_number_uav}, energy consumption increases almost linearly with the number of UAVs. It can be observed that when the number of UAVs is small, our proposed approach consumes slightly more energy to meet heterogeneous coverage requirements. However, as the number of UAVs increases, other methods struggle to balance coverage and energy efficiency. When $N_{\mathcal{K}}=10$, our approach achieves $7\sim 16\%$ lower energy consumption than ISAC-S2DC, demonstrating its energy-saving capability in HetUAVNs.

Table~\ref{Tab:runtime_diff_uav} shows how the runtime of the proposed algorithm varies with the number of UAVs. As shown in the table, the inference latency of LLM-HeMARL remains ultra-low and stable, fluctuating between $15$ and $16$ ms regardless of UAV swarm size. In contrast, the S2DC algorithm exhibits polynomial growth in runtime due to the computational cost of handling complex coupled constraints within the optimization problem. In contrast to SCA-S2DC, which incurs an average runtime per time slot on the order of tens of seconds to minutes, our algorithmic framework significantly alleviates the computational burden by decoupling trajectory design from precoding. Moreover, with ongoing advances in computing hardware, the runtime is expected to decrease further.

From the above analysis, we can conclude that as the number of UAVs increases and the number of decision variables increases, the advantages of the approach based on the hierarchical solution framework (LLM-HeMARL-S2DC, ISAC-S2DC and MASAC-S2DC) become increasingly evident. Furthermore, due to limited global experience sharing and lack of expert guidance, ISAC-based methods may underperform compared to MASAC-based counterparts in certain scenarios under high randomness. Fortunately, the integration of LLM-derived expert policy compensates for these limitations, enabling superior overall performance in HetUAVN environments.

\begin{table}[t]
	\caption{\textsc{Runtimes Scales (ms).}}
	\renewcommand{\arraystretch}{1.5}
	\label{Tab:runtime_diff_uav}
	\begin{tabular}{|c|c|c|c|c|c|}
		\hline
		\diagbox{\textbf{Algo.}}{\textbf{$N_\mathcal{K}$}}       & 2      & 4       & 6       & 8       & 10      \\ \hline
		S2DC       & 425.34 & 1135.77 & 2581.45 & 4071.89 & 5916.36 \\ \hline
		LLM-HeMARL & 15.43  & 15.27   & 16.34   & 16.84   & 16.12   \\ \hline
	\end{tabular}
\end{table}

\section{Conclusion}
This paper has considered more practical scenarios and explored the trade-off between network security and energy consumption of HetUAVNs for the first time. We have analyzed the unique challenges in secure HetUAVNs and modeled the underlying problems using a multi-objective framework. To handle the high coupling and non-convex complexity, we have proposed a hierarchical optimization framework, in which we have applied the S2DC algorithm in the inner layer and the LLM-HeMARL algorithm in the outer layer to jointly optimize precoding and trajectory to maximize the secrecy rate and minimize the energy consumption. Simulation results have demonstrated that the proposed hierarchical optimization framework effectively decouples the complex joint optimization problem, leading to substantial improvements in system performance. Compared to conventional RL baselines, the integration of LLM-generated expert policies enables UAV agents to make heterogeneity-aware decisions, resulting in significant performance gains in terms of convergence speed and solution quality. Moreover, the robustness and scalability of the proposed approach were validated under different random seeds and UAV swarm sizes.

\bibliographystyle{IEEEtran}
\bibliography{IEEEabrv, ref}

\vfill

\end{document}